\documentclass[twocolumn]{aastex62}

\usepackage{subfigure}
\usepackage{url}
\usepackage{hyperref}
\usepackage{longtable}
\usepackage{natbib}
\usepackage{amsmath}
\usepackage{listings}
\usepackage[normalem]{ulem}
\usepackage{bm}
\usepackage{comment}
\usepackage{array}
\newcolumntype{P}[1]{>{\centering\arraybackslash}p{#1}}
\newcolumntype{M}[1]{>{\centering\arraybackslash}m{#1}}

\usepackage{xcolor, fontawesome}
\definecolor{twitterblue}{RGB}{64,153,255}

\newcommand{\twitter}[1]{\href{https://twitter.com/#1 }{\textcolor{twitterblue}{\faTwitter}\,\tt \textcolor{twitterblue}{@#1}}}

\newcommand{\github}[1]{\href{https://github.com/#1 }{\textcolor{black}{\faGithub}\,\tt \textcolor{black}{@#1}}}

\definecolor{Code}{rgb}{0,0,0}
\definecolor{Decorators}{rgb}{0.5,0.5,0.5}
\definecolor{Numbers}{rgb}{0.5,0,0}
\definecolor{MatchingBrackets}{rgb}{0.25,0.5,0.5}
\definecolor{Keywords}{rgb}{1,0,0}
\definecolor{self}{rgb}{0,0,0}
\definecolor{Strings}{rgb}{0,0.63,0}
\definecolor{Comments}{rgb}{0,0.63,1}
\definecolor{Backquotes}{rgb}{0,0,0}
\definecolor{Classname}{rgb}{0,0,0}
\definecolor{FunctionName}{rgb}{0,0,0}
\definecolor{Operators}{rgb}{0,0,0}
\definecolor{Background}{rgb}{0.98,0.98,0.98}
\definecolor{Booleans}{rgb}{0.572,0,0.572}
\definecolor{BuiltinFunction}{rgb}{0.572,0,0.572}
\definecolor{BuiltinConstant}{rgb}{0.572,0,0.572}
\definecolor{Asterisk}{rgb}{0.670,0,1}

\lstdefinelanguage{Python}{
    	numbers=left,
    	numberstyle=\footnotesize,
    	numbersep=7pt,
    	xleftmargin=1.26em,
    	framextopmargin=2em,
    	framexbottommargin=2em,
    	showspaces=false,
    	showtabs=false,
    	showstringspaces=false,
    	frame=l,
    	tabsize=4,
    	stepnumber=1,
	basicstyle=\small\ttfamily,
    	backgroundcolor=\color{Background},
	stringstyle=\ttfamily\color{Strings},
	morekeywords={import,from,class,def,while,if,in,elif,else,not,or,print,break,continue,return,access,as,except,exec,finally,global,import,lambda,pass,print,raise,try,assert},
    	keywordstyle={\color{Keywords}\bfseries}, 
	otherkeywords={[2]*},
	keywordstyle={[2]\color{Asterisk}},
}

\usepackage{color}

\newcommand{\shrug}{\texttt{\raisebox{0.75em}{\char`\_}\char`\\\char`\_\kern-0.5ex(\kern-0.25ex\raisebox{0.25ex}{\rotatebox{45}{\raisebox{-.75ex}"\kern-1.5ex\rotatebox{-90})}}\kern-0.5ex)\kern-0.5ex\char`\_/\raisebox{0.75em}{\char`\_}}}

\newcommand{\kep}{{\it Kepler}}
\newcommand{\ktwo}{{\it K2}}
\newcommand{\tess}{{\it TESS}}

\newcommand{\Gaia}{{\it Gaia}}

\newcommand{\stella}{\texttt{stella}}
\newcommand{\banyan}{\texttt{BANYAN-$\Sigma$}}

\newcommand{\lightkurve}{\texttt{lightkurve}}

\newcommand{\chicago}{Department of Astronomy and Astrophysics, University of
Chicago, 5640 S. Ellis Ave, Chicago, IL 60637, USA}

\newcommand{\nsf}{NSF Graduate Research Fellow}

\newcommand{\unsw}{School of Physics, University of New South Wales, Sydney, NSW 2052, Australia}

\newcommand{\flatiron}{Flatiron Institute, Simons Foundation, 162 Fifth Ave, New York, NY 10010, USA}

\newcommand{\fermi}{Fermi National Accelerator Laboratory, P.O. Box 500, Batavia, IL 60510, USA}

\newcommand{\kavli}{Kavli Institute for Cosmological Physics, University of Chicago, Chicago, IL 60637, USA}

\newcommand{\masstech}{Department of Physics, and Kavli Institute for Astrophysics and Space Research, Massachusetts Institute of Technology, Cambridge, MA 02139, USA}

\newcommand{\gsfc}{Exoplanets and Stellar Astrophysics Laboratory, Code 667, NASA Goddard Space Flight Center, Greenbelt, MD 20771}

\newcommand{\austin}{Department of Astronomy, The University of Texas at Austin, 2515 Speedway Boulevard, Austin, TX 78712, USA}

\begin{document}
\title{Flare Statistics for Young Stars from a
Convolutional Neural Network Analysis of \tess\ Data}

\shorttitle{Young stellar activity} 
\shortauthors{Feinstein et al.}

\author[0000-0002-9464-8101]{Adina~D.~Feinstein}
\altaffiliation{\nsf}
\affiliation{\chicago}

\author[0000-0001-7516-8308]{Benjamin~T.~Montet}
\affiliation{\unsw}

\author[0000-0003-4142-9842]{Megan~Ansdell}
\affiliation{\flatiron}

\author[0000-0001-6706-8972]{Brian~Nord}
\affiliation{\chicago}
\affiliation{\fermi}
\affiliation{\kavli}

\author[0000-0003-4733-6532]{Jacob~L. Bean}
\affiliation{\chicago}

\author[0000-0002-3164-9086]{Maximilian~N.~G{\"u}nther}
\altaffiliation{Juan Carlos Torres Fellow}
\affiliation{\masstech}

\author[0000-0002-4020-3457]{Michael~A.~Gully-Santiago}
\affiliation{\austin}

\author[0000-0001-5347-7062]{Joshua~E. Schlieder}
\affiliation{\gsfc}

\correspondingauthor{Adina~D.~Feinstein;\\ \twitter{afeinstein20}; \github{afeinstein20}} \email{afeinstein@uchicago.edu}


\begin{abstract}

All-sky photometric time-series missions have allowed for the monitoring of thousands of young ($t_{\rm age} < 800$\,Myr) to understand the evolution of stellar activity. Here we developed a convolutional neural network (CNN), \stella, specifically trained to find flares in \textit{Transiting Exoplanet Survey Satellite} (\tess) short-cadence data. We applied the network to 3200 young stars to evaluate flare rates as a function of age and spectral type. The CNN takes a few seconds to identify flares on a single light curve. We also measured rotation periods for 1500 of our targets and find that flares of all amplitudes are present across all spot phases, suggesting high spot coverage across the entire surface. Additionally, flare rates and amplitudes decrease for stars $t_{\rm age} > 50$\,Myr across all temperatures $T_{\rm eff} \geq 4000$\,K, while stars from $2300 \leq T_{\rm eff} < 4000$\,K show no evolution across 800 Myr. Stars of $T_{\rm eff} \leq 4000$\,K also show higher flare rates and amplitudes across all ages. We investigate the effects of high flare rates on photoevaporative atmospheric mass loss for young planets. In the presence of flares, planets lose 4-7\% more atmosphere over the first 1 Gyr. \stella\ is an open-source Python tool-kit hosted on GitHub and PyPI.

\end{abstract}

\keywords{PMS stars, Stellar activity, Stellar rotation, Convolutional neural networks, Time series analysis}

\section{Introduction} \label{sec:intro}


Young stars are appealing targets for studying the early stages of stellar and exoplanet evolution. Stellar flares are energetic events caused by the reconnection of magnetic field lines and there is evidence that they abound on very active coronal sources, such as pre-main sequence stars \citep{benz10}. Flare rates are of particular interest due to their effects on the early stages of exoplanet evolution. They have been seen to cause phenomena such as increased photoevaporation of inner protoplanetary disks \citep{benz10} and irreparable chemical changes to exoplanet atmospheres \citep{venot16}. Flares can also expedite atmospheric erosion \citep{Lammer07}, particularly over the first few hundred million years, when the atmospheres of young planets are still contracting \citep{owen17}.

The advent of high-precision time-series photometric missions, e.g. \kep/\ktwo\ \citep{Borucki10} and the \textit{Transiting Exoplanet Survey Satellite} \citep[\tess;][]{Ricker14}, have allowed for detailed studies of stellar activity across both spectral type and age \citep[e.g.][]{davenport14, Hawley14}. Using data from the 4-year \kep\ mission, \cite{Davenport16} identified $\sim 850,000$ flares across 4000 stars spanning G0 - M4 spectral types and found a potential flare saturation limit as a function of Rossby number. It was later noted this sample may have significant pollution due to variable stars in this sample \citep{Davenport19}. More recently, \cite{guenther19_flares} conducted a flare search of all two-minute cadence stars observed in Sectors 1 and 2 of \tess. Of the entire sample, they found 1228 flaring stars with a total of $\sim 8700$ flares, with the most flares occurring on $>$ M4.5 dwarfs.


Stars are known to spin-down as they age \citep{Soderblom10}. The generation of magnetic fields is driven by the conversion of kinetic to magnetic energy through dynamo theory. \cite{notsu13} and \cite{candelaresi14} found flare rates decrease with rotation period, a proxy for age \citep{Noyes84}, and flare energy. Additionally, \cite{notsu13} found that the magnetic energy released by superflares identified in \kep\ data can be explained by the magnetic energy stored near starspots, concentrated regions of magnetic field lines \citep{dun07}. More recently, \cite{ilin19} identified and analyzed \ktwo\ data for the Pleiades, Praesepe, and M67 and found a decrease in flaring activity with an increase in age. There was a sharper decrease in activity for stars with higher effective temperatures, $T_{\rm eff}$. 

Using short-cadence \ktwo\ light curves, \cite{doyle18} explored the relation between starspots and flares. The identified flares appeared to show no correlation with the phase of the spot grouping, i.e. the spottier hemisphere had similar flare counts to the less spotty hemisphere. \cite{doyle18} also noticed that faster rotators showed greater flaring activity. Repeating similar studies, \cite{doyle19} looked at 167 M dwarfs and \cite{doyle20} looked at 158 G-M main sequence stars in \tess\ short-cadence light curves observed in Sectors 1-13. Again, they found no correlation between the spot phase and flares at any energy.

\cite{roettenbacher18} conducted a similar study with a sample of 119 main-sequence stars observed during the \kep\ mission. Within this sample of stars, \cite{roettenbacher18} found that low-amplitude flares are correlated with spot groupings. However, the superflares followed similar trends and were detected across all spot phases. Detailed studies of the Sun have revealed that low-amplitude flares are associated with the presence of sunspots \citep{zhang08}. While \cite{roettenbacher18} found similar trends to that of the Sun, \cite{doyle18, doyle19, doyle20} did not.

The previously discussed flare detection studies relied on detrending a light curve and using outlier detection heuristics for identifying flare events. Most detection algorithms apply similar statistical criteria to those defined in \cite{chang15} on a detrended light curve. The criteria defined in \cite{chang15} predominantly rely on at least three consecutive cadences that lie $\geq 3\sigma$ above the median of the light curve. \cite{guenther19_flares} relied on fitting a spline to the underlying stellar variability and identifying flare candidates as at least six minutes (three cadences) of flux $3\sigma$ above the detrended light curve flux. \cite{doyle18, doyle19, doyle20} removed the underlying spot modulation and marking at least two consecutive cadences $\geq 2.5\sigma$ above the median light curve as a potential flare. \cite{roettenbacher18} used a RANdom SAmple Consensus (RANSAC) algorithm, which identifies and subtracts inliers (the underlying light curve) before searching for outliers above a given detection limit \citep{vida18}. The outliers are flagged as flare candidates. 

However, light curve detrending has the potential to remove low-energy flares because aggressive spot variability models can include cadences that may belong to these flares. In turn, this could bias flare detection towards the highest energy flares and affect modeled flare energies. For example, \cite{davenport14} do not find flares below $\sim 10^{33}$ ergs across a range of spectral types. This is additionally true for having a set outlier cut, where low-energy flares may not lie above this threshold. Because individual flares are believed to originate from the same physical processes, they exhibit similar time evolution \citep{benz10} and therefore similar characteristic shapes in photometric data. Although at some point there will be a noise level to the flare amplitudes we are able to detect, this proves promising for using machine learning algorithms to identify such a feature.


Machine learning algorithms comprise a set of techniques in which models are derived primarily from the data presented to them. Generically, the key informative or distinguishing features of these data must also be presented to the algorithms. However, in a subset of techniques, deep learning algorithms can be presented with raw data, and the models themselves identify the critical informative features. 

Several instances of machine learning have already been applied to time series data for planetary and stellar studies. \cite{shallue18} trained CNNs to identify exoplanet transits and present discoveries of new planet candidates. \cite{ansdell18} expanded upon these techniques by including centroid positions of the stars to train their CNN. All three methods relied on detrended light curves and hand-labeled transit events. \cite{pearson18} also used CNNs to search for new exoplanet candidates and trained their network on simulated data with underlying stellar variability. They trained on light curves with underlying photometric variability. 

As young stars are more rapidly rotating than main-sequence stars, it may be expected that these stars exhibit higher flare rates. Due to the high magnetic activity and spot coverage of rapidly rotating stars \citep{Montet17, morris2020}, pre-main sequence (PMS) and young ($t_{\rm age} < 800$ Myr) stars may prove to be excellent targets for understanding the relationship between starspots and flares. Here, we present the results of a CNN trained to identify flares in short-cadence \tess\ data as applied to young stars observed in Sectors 1-20 of \tess. The CNN is part of the new open-source software package, \stella\footnote{JOSS submitted; \url{https://joss.theoj.org/papers/70892fb54a060e00c5707d0111b13e06}}. Our analysis is aimed to better understand flare statistics down to low energies, how they relate to spot groupings, and general trends in flare rates with respect to age and spectral type. 

The paper is presented as follows: Section~\ref{sec:cnn} will discuss the details of creating and validating the CNN. Section~\ref{sec:analysis} will describe the young stellar sample, rotation period measurements, and flare identification used in this work. Section~\ref{sec:results} will present our analysis of the newly identified flares with relation to spots and Section~\ref{sec:discussion} will provide a discussion on these results and implications for exoplanets. We will conclude in Section~\ref{sec:conclusion}.

\section{The Convolutional Neural Network: \stella}\label{sec:cnn}


Neural networks learn features from a training data set that consists of input examples with corresponding classifying labels. When presented with samples labeled by a given class, supervised learning techniques are used to optimize the network weights such that they collectively identify sets of features relevant for those classes. Neural networks are often referred to as ``deep learning" because they consist of a series of hierarchical computational layers, where each layer is made up of a series of neurons, or scalar valued units \citep{aggarwal14}. The input to a neuron is calculated by weighting the outputs from neurons in the previous layer, then adding a bias term and applying a non-linear activation function \citep{rosenblatt61}. The activation function is a key component of a neural network because it enables the network to solve non-linear problems. For classification problems (such as used in this work), the final output of a neural network is normalized to a value between 0 and 1, which represents a ranking (but not necessarily a probability, see Section~\ref{subsec:probs}) of the input example belonging to the positive class (in our case, flares). 

Fully connected neural networks are not designed to explicitly account for spatial structure in data. Neurons in a Convolutional Neural Network (CNN), in contrast, are only connected to local regions in the preceding layer, allowing them to detect local features in the data. CNNs are promising for astronomical purposes as they can be optimized for computer vision tasks. This means the algorithms can identify patterns and features within complex data sets without requiring prior information on said features. Stellar flares exhibit characteristic shapes in time and flux space and therefore training a CNN to identify this specific feature is promising. In CNNs, a convolution layer applies filters to the data, resulting in feature maps from each layer \citep{lecun15}. The weights in a given filter increase when they see certain types of features in the data that they have been optimized for, and this information is then fed into the next layer. The number of feature maps, as well as the complexity of the detected features, typically increases with network depth. Pooling layers aggregate values within small neighborhoods (strides) of neurons along the input by calculating and outputting a summary statistic (e.g., mean, median, max) \citep{krizhevsky12}. Pooling is primarily used to reduce the number of input pixels to the next layer. This allows the feature maps in the next convolutional layer to have a wider view of the input but at lower resolution and without increasing the number of trainable parameters.

Neural networks are useful for scientific research for several reasons. First, they can in many cases outperform physically parameterized modeling or human inspection techniques in their prediction accuracy: this is attributed to their flexibility in identifying critical data representations that are indicative of important aspects of the data with which they are presented. Moreover, unlike human vetting, deep learning models are systematic, which is important for calculating metrics like occurrence rates. Neural networks are also typically fast in the inference stage: once models are trained, it can take just seconds to apply to hundreds of thousands of new examples; similarly, they are sometimes upgradable, such that improved models can be quickly re-trained and applied to new data. However, networks can suffer from challenges with bias and uncertainty quantification. The data sets used to train a network contain biases, which the network is optimized to represent. Additionally, there are an increasing number of methods for quantifying errors in networks, but it is an area of active study to provide accurate and interpretable error estimates.

\subsection{Data preparation: Training, Validation, and Test Sets}\label{subsec:cnn-data}

The \tess\ targets used in training, validating, and testing the CNN are taken from \cite{guenther19_flares}, who searched for flares in the first two sectors of the \tess\ mission. The light curves consist of integrated flux measurements taken at two minute cadence over roughly 27 days; they were made publicly available with the first \tess\ data releases through the Mikulski Archive for Space Telescopes (MAST). Similarly to \cite{guenther19_flares}, we split each light curve into individual orbits, and normalized the Simple Aperture Photometry flux (SAP flux) separately for each orbit. 

For supervised learning tasks, neural networks require input data that are uniformly sampled to train properly. For the inputs to the CNN implemented here, we used a data set of one-dimensional time series where all elements have the same number of 2-minute cadences. We found that a length of 200 cadences provided enough information about the baseline flux surrounding a given flare. Longer baselines often predicted high probabilities for both rotational signatures and flares instead of just flares. This baseline also provided ample flare and non-flare sets to train, validate, and test on. Following the methods of \cite{pearson18}, we ensured all known flare peak times from the \cite{guenther19_flares} catalog were centered at the 100\textsuperscript{th} cadence (i.e. centered). Each of these light curve snippets are hereafter referred to as a ``sample." All of the discussed steps (e.g. training and ensembling a series of CNN models) in this section are incorporated into the open-source Python package \stella.\footnote{\url{https://github.com/afeinstein20/stella}} \stella\, and the CNN architecture described here, is specifically tailored for finding flares in \tess\ short-cadence light curves and should not be applied to other photometric time-series data.

\subsection{Labels}\label{subsec:cnn-labels}

\begin{figure}[!ht]
\begin{center}
\includegraphics[width=0.45\textwidth,trim={0.25cm 0 0 0}]{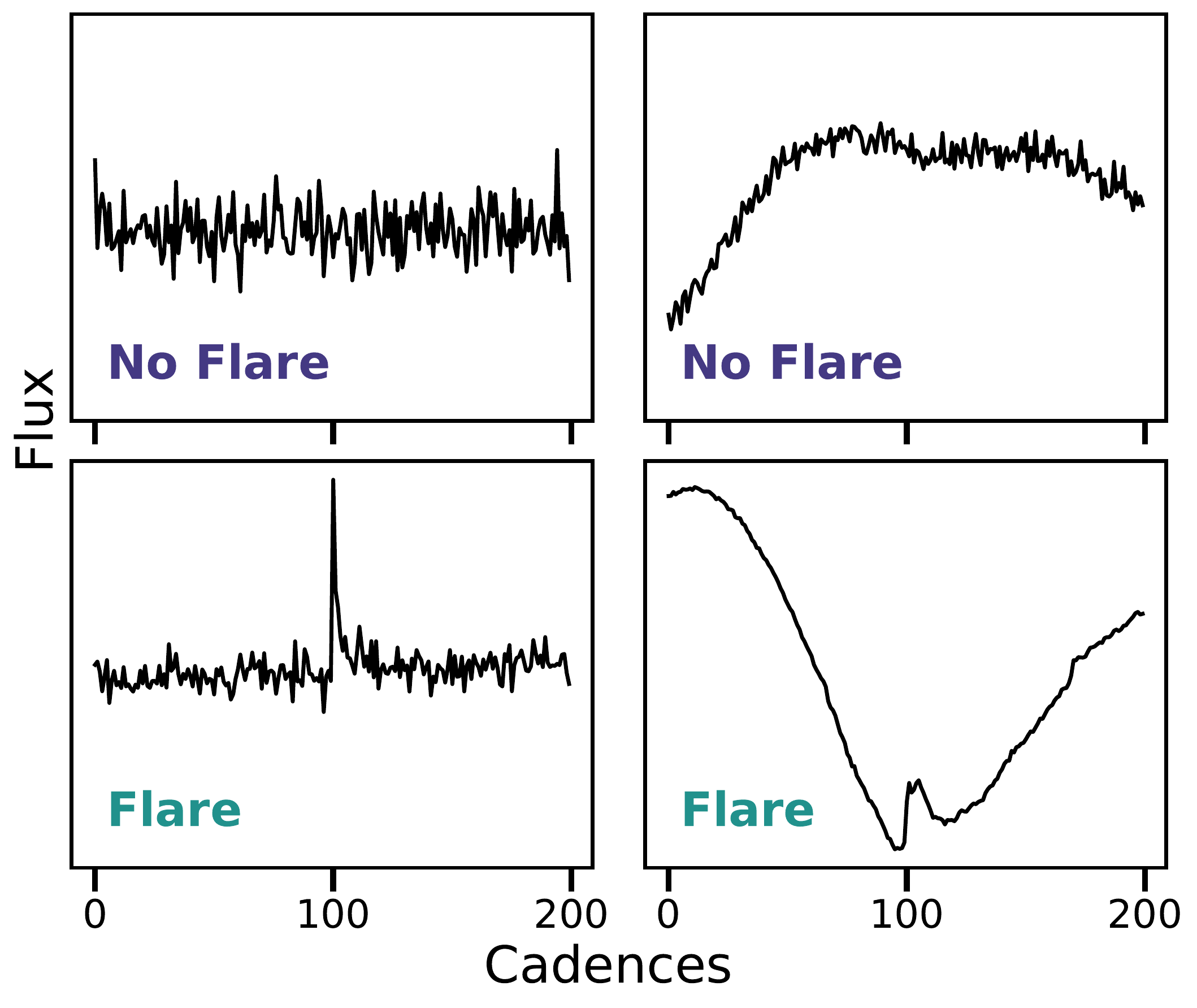}
\caption{Samples in the training set. Using flares identified in \cite{guenther19_flares}, we created a training set of non-flares (top) and flares (bottom), each of equal 200 cadence length. The light curves were not normalized. We include within the non-flare cases some examples of obvious spot modulation (upper right) so the CNN will ignore this variability and focus on the characteristic flare shape.} \label{fig:training_set}
\end{center}
\end{figure}

We used a binary labeling scheme of ``flare" and ``non-flare" for the samples (see Figure~\ref{fig:training_set} for examples of the samples). For the flare examples, we used the peak times of flares identified by \cite{guenther19_flares}. Non-flare samples were centered on locations in the light curves at least 100 cadences from a flare. Our final training set contains 5389 hand-labeled flare examples and created 17684 non-flare examples for a 30\% positive class data set. We then randomly divided the data set into training (80\%), validation (10\%), and test (10\%) sets. We used the validation set to tune the network and train hyperparameters and then used the test set to evaluate the final model performance (Section~\ref{subsec:cnn-results}).

Due to the detection limitations of the flare identifying methods used in creating the training set, we encountered a few issues. First, not all flares, particularly those at low-energy, were identified in the original catalog and therefore have a ``non-flare" label in the training set (Figure~\ref{fig:confusion}; false negatives). Second, we found the catalog is off in peak flare time for some cases and therefore have been classified as false positives when evaluating the validation set. This is because the flare was not at the center cadence of the example.

\subsection{Network Architecture \& Training}\label{subsec:cnn-model}

\begin{figure}[!ht]
\begin{center}
\includegraphics[width=0.46\textwidth,trim={0.25cm 0 0 0}]{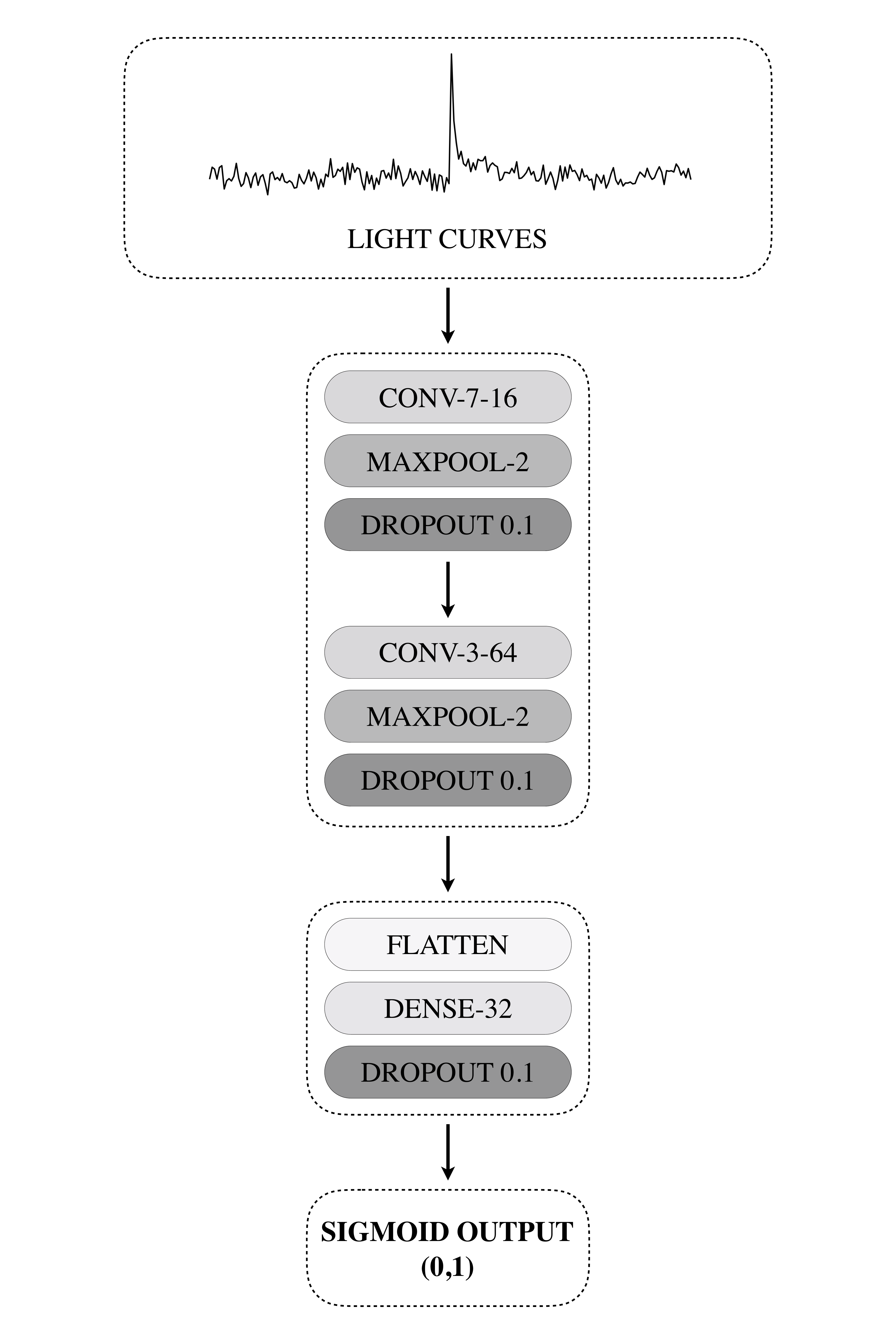}
\caption{The architecture of the \stella\ CNN. The training set consists of ``flare" and ``non-flare" cases, where flares are in the center of a 200-cadence section of the light curve. ``CONV-$<$kernel size$>$-$<$number of filters$>$": a 1D convolutional layer with affiliated parameters. ``MAXPOOL-$<$pool size$>$": 1D max pooling tensor. ``DROPOUT-$<$dropout fraction$>$": Drops out fraction of input units to prevent over-fitting. ``DENSE-$<$units$>$": Creates densely-connected layers of specified units. ``SIGMOID OUTPUT": ``score" of being part of the positive class. } \label{fig:schematic}
\end{center}
\end{figure}

Our CNN architecture, shown in Figure~\ref{fig:schematic}, is implemented in \texttt{tf.keras}, which is TensorFlow's \citep{Abadi2016} open source, high-level implementation of the Keras API specification \citep{keras18}. The network consists of a one-dimensional convolutional column with global max pooling and dropout, the results of which are flattened and fed into a series of fully connected (or ``dense") layers ending in a sigmoid function that produces an output in the range [0,1]. This output loosely represents the ``score" of how likely a given example is a flare (1) or non-flare (0) event. 

The usage of global max pooling and dropout are standard practices with state-of-the-art CNNs \citep[e.g., ImageNet;][]{He2016} to improve model performance. Max pooling downsamples the feature maps in a given layer by ``pooling" (taking the maximum of) the output neurons within a given region, which can reduce the number of model parameters while increasing generalization \citep[e.g.,][]{Lin2013}. Dropout helps prevent model over-fitting by randomly ``dropping'' (or setting to zero) some fraction of the output neurons in a given layer during training to prevent the model from becoming overly dependent on any of its features \citep{Srivastava2014}. 

Training neural networks involves inputting samples and then minimizing a cost function that measures how far off the network's predictions are from the truth. This is done through back propagation, which updates the model parameters to reduce the value of the cost function. For model training, we used the Adam optimization algorithm \citep{Kingma2014} to minimize the binary cross-entropy error function. The Adam optimizer was run with a learning rate of $\alpha=10^{-3}$ (this controls the degree to which the weights are updated with each iteration), exponential decay rates of $\beta=0.9$ and $\beta=0.999$ (for the first and second moment estimates), and $\epsilon=10^{-8}$ (a small number to prevent any division by zero in the implementation).

\subsection{Model Evaluation}\label{subsec:cnn-results}

The exact model architecture, kernel sizes, etc. were chosen based on a trial and error approach to avoid over-fitting the model. Over-fitting was evaluated using four standard machine learning metrics: accuracy, precision, recall, and average precision. Accuracy is the fraction of correct classifications by the model for both classes (flares and non-flares), at a given threshold for deciding when the model output becomes a positive class; we use a threshold of 0.5 for our accuracy calculations and overall analysis. Precision is the fraction of flares classified as flares that are true flares, while recall is the fraction of true flares recovered by the model. The average precision summarizes the precision-recall curve as the weighted mean of precisions achieved at each threshold, i.e. it is the area under the precision-recall curve evaluated across various threshold values, not just 0.5, as the precision is.

\begin{figure}[!ht]
\begin{center}
\includegraphics[width=0.45\textwidth,trim={0.25cm 0 0 0}]{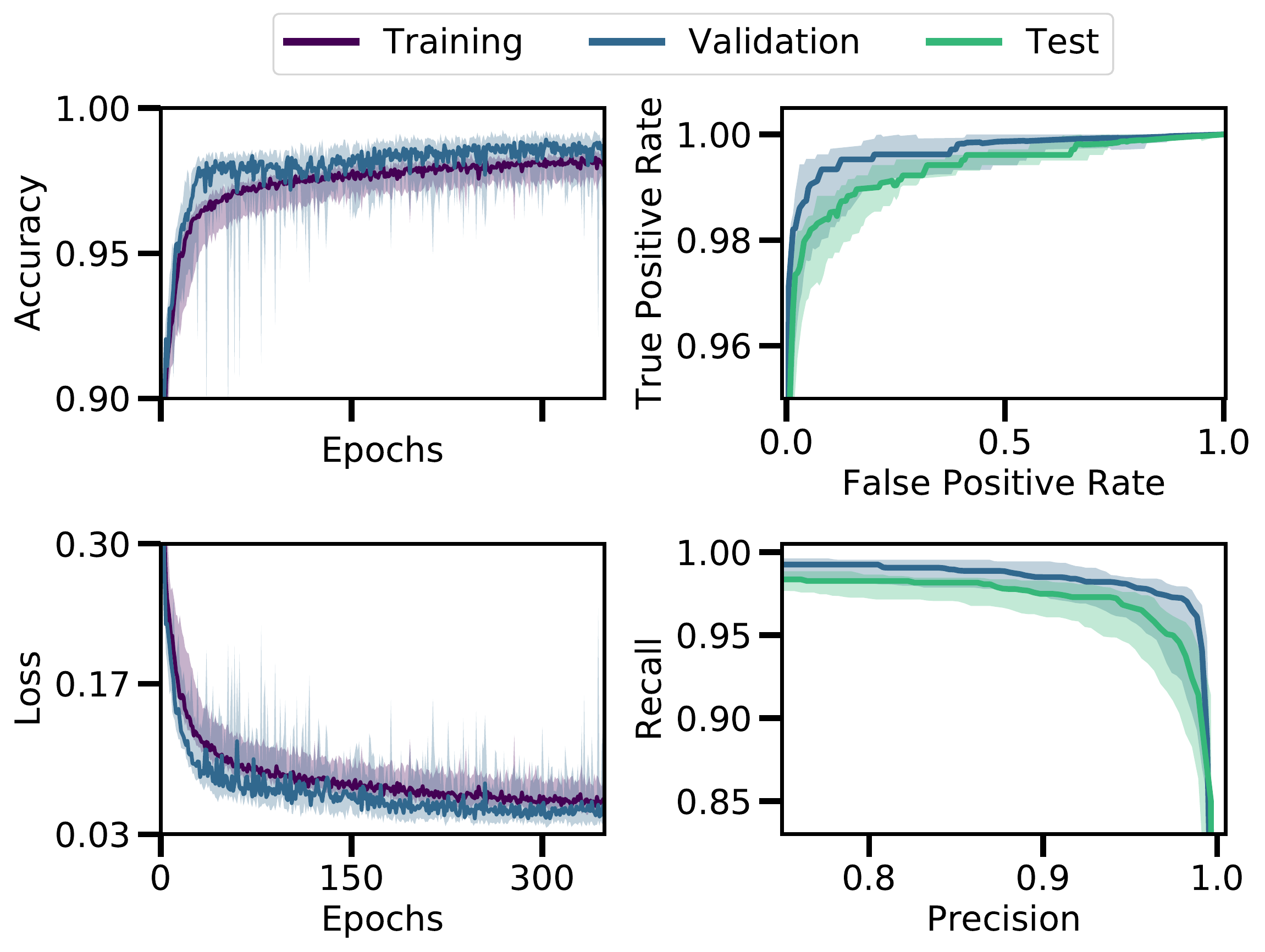}
\caption{The results of training 10 \stella\ models demonstrated through several standard metrics. Each model was initialized with a different random seed. The left column demonstrates the evolution of the accuracy (top) and loss (bottom) functions for the training and validation sets over the number of epochs trained on. The right column evaluates the performance of \stella\ on the validation and test sets through the receiver operating characteristics (ROC; top) and the precision-recall curve (bottom). The line represents the median curve. The shaded regions represent the 5\textsuperscript{th} and 95\textsuperscript{th} percentiles across the 10 models.} \label{fig:loss-acc}
\end{center}
\end{figure}

We use model ensembling to produce a distribution of 10 independently trained versions of the same model with different random parameter initializations. We then average the predictions across the 10 models and evaluate the above mentioned metrics to establish more robust classifications. Diagnostic plots for the performance of the 10 \stella\ models are shown in Figure~\ref{fig:loss-acc}. The left column demonstrates the evolution of the accuracy (top) and loss (bottom) functions for both the training and validation sets. As the accuracy converges towards 1 and the loss towards 0, we note that the models are not being over-fit to the data. The receiver operating characteristics (ROC; top right) visualizes the trade-off of true- and false-positive rates identified in the validation and test sets. The ROC is summarized by the area under the curve (AUC), where an AUC $\sim 1$ reveals the classification was successful. We obtain an AUC of $0.997 \pm 0.002$ and $0.993 \pm 0.002$ for the validation and test sets, respectively. The precision-recall curves for the validation and test sets are shown in the bottom right of Figure~\ref{fig:loss-acc}.

\begin{figure}[!ht]
\begin{center}
\includegraphics[width=0.45\textwidth,trim={0.25cm 0 0 0}]{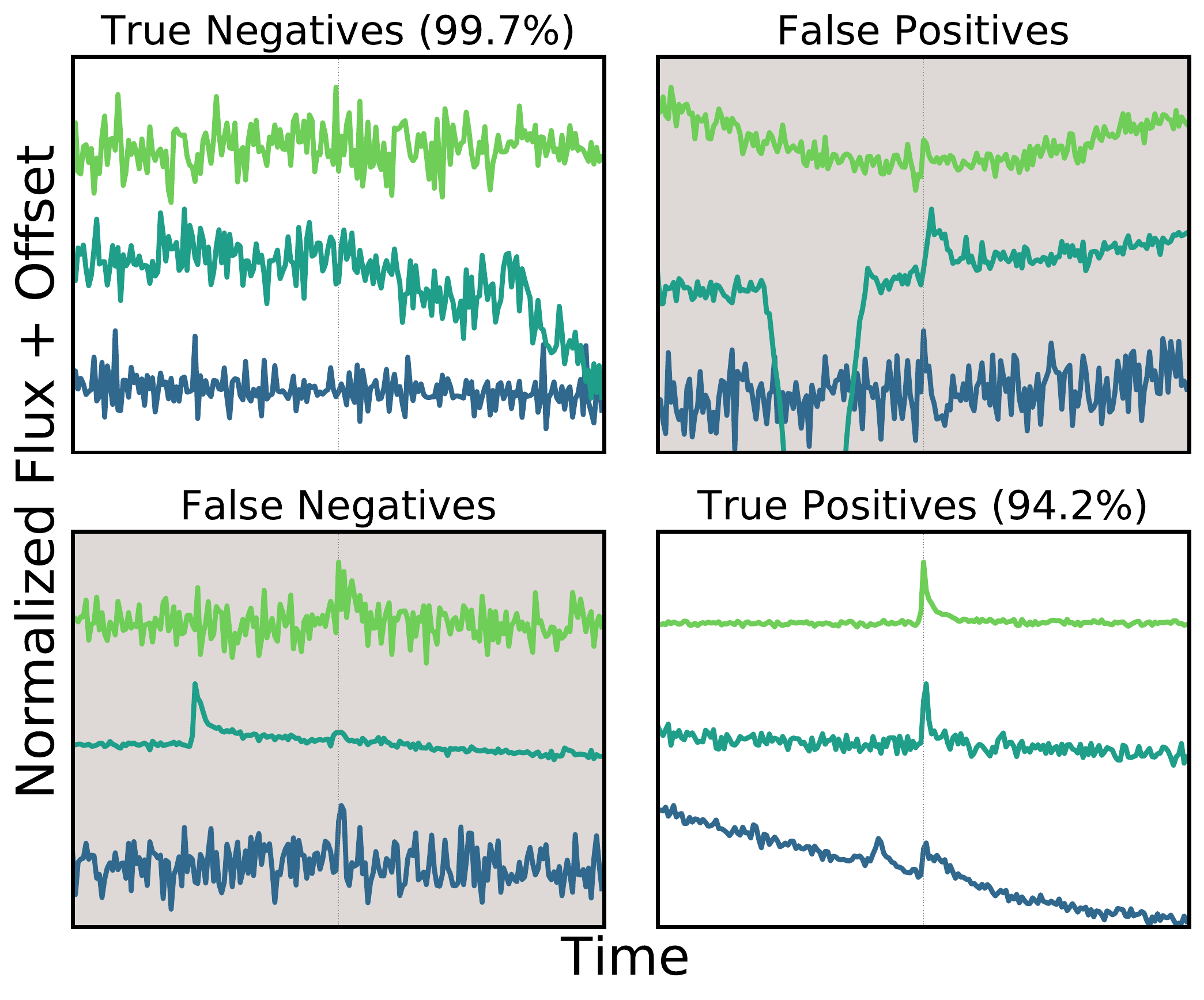}
\caption{The confusion matrix for the \stella\ test set. The vertical gray dashed line shows the location of the flare. These samples have been re-scaled so the relative sizes are meaningless; they are offset for clarity. The flares identified as false negatives tend to have flare shapes that deviate from typical positive samples and are probably under-represented in the training set. False positives have flare-like shapes, and could either be noise structures identified as flares or true flares which were unlabeled in the test set due to limitations of previous flare-identifying techniques. The percentages represent the percent of that class recovered in the validation set.} \label{fig:confusion}
\end{center}
\end{figure}

Ensembling makes comparisons between different model architectures more robust because it averages over the stochastic differences in the individual models due to their different random parameter initializations. Moreover, ensembling improves model performance because the individual models can perform slightly better (or worse) in different regions of input space, in particular when the training set is small and thus prone to over-fitting. The ensembled results for the test set are: 0.9844 (accuracy), 0.9878 (precision), and 0.9419 (recall). The average precision is 0.9923. We show the confusion matrix with examples of typical light curves in each category in Figure~\ref{fig:confusion}.

\begin{figure}[!ht]
\begin{center}
\includegraphics[width=0.45\textwidth,trim={0.25cm 0 0 0}]{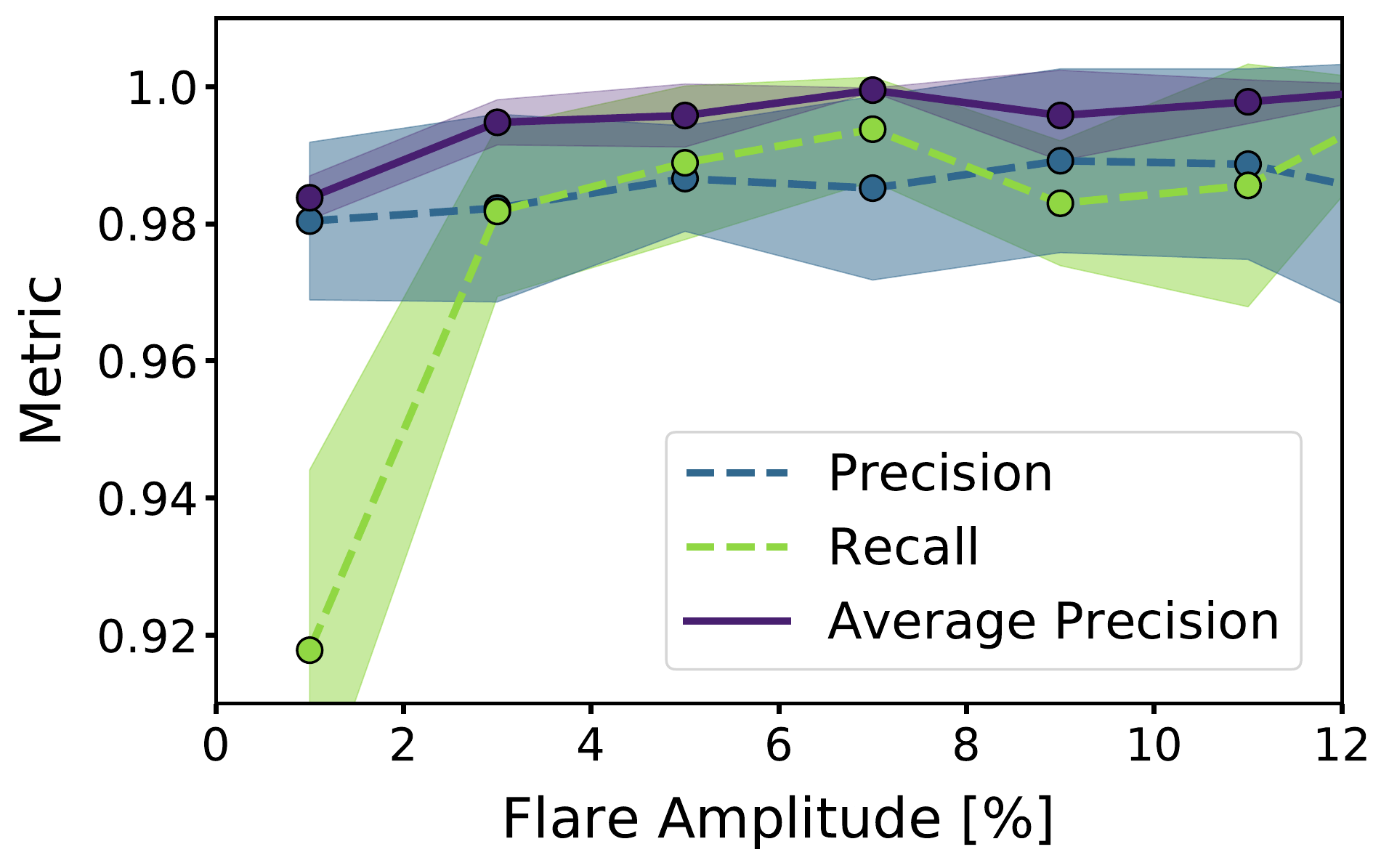}
\caption{The results of how each evaluation metric (legend) changes with flare amplitude. The lines are the averages of the cross-validation k-fold results, while the shaded regions show their standard deviations in each amplitude bin. We use a threshold of 0.5 to calculate the recall (dashed green line), while the average precision (solid purple line) does not require choosing a threshold value.} \label{fig:amp_metrics}
\end{center}
\end{figure}

Additionally, we run a k-fold cross-validation on the combined training and validation sets. Cross-validation is a method of evaluating model generalization performance that is more robust than using training and test sets; in k-folds cross-validation, the data are instead split repeatedly into k parts of equal size and multiple models are trained (here we use k = 5). We show the results of this in Figure~\ref{fig:amp_metrics} to illustrate how the model performance changes with flare amplitude; as expected, smaller flares are more difficult to classify. 

\subsection{Determining Probabilities of Events}\label{subsec:probs}

\begin{figure}[!ht]
\begin{center}
\includegraphics[width=0.4\textwidth,trim={0.25cm 0 0 0}]{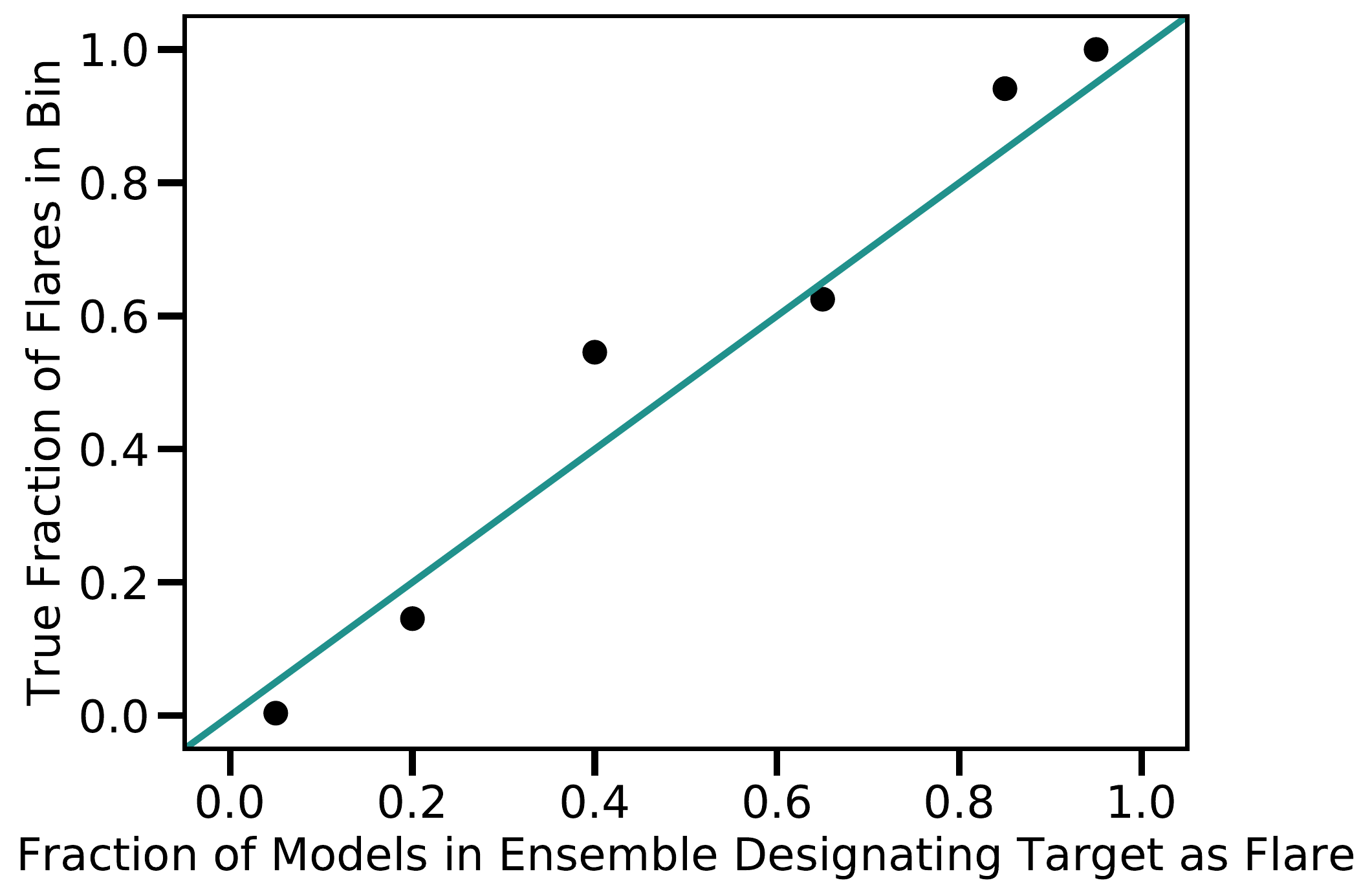}
\caption{The output of the CNN is not a probability and needs to be calibrated to such. Here, we demonstrate that the output value of the CNN (x-axis) corresponds very well to the true fraction of flares (y-axis) and thus no calibration needs to be completed. The output of the \stella\ CNN can be taken as a true ``probability." A one-to-one line is plotted in teal.} \label{fig:calibration}
\end{center}
\end{figure}

The output of a neural network classifier is in general not a probability \citep{NiculescuMizil05}, but rather a ranking or score of a given example being associated with the positive class (in our case, a true flare). In some cases, however, the network output may happen to be calibrated such that it is a good estimate of the probability. To test whether this is the case, we calculated the fraction of models across our ensemble that return the flare classification (assuming a threshold of 0.5) for each example in the test set. We then made a histogram of these values with bins spaced such that there were the same number of targets in each bin. For each bin, we then calculated the fraction of true flares using the known labels and compared this value to the fraction of flare classifications by the model: if the model output happens to be calibrated, then these two values should match in each bin. The results are shown in Figure~\ref{fig:calibration}, which illustrates a monotonically increasing function near the one-to-one line, indicating that the model is indeed fairly well calibrated and we can use our model output as a rough estimate of probability. We note that there is no measure of uncertainty associated with each classification.

\section{Analysis}\label{sec:analysis}

We applied the trained CNN to young stars observed at \tess\ short-cadence for flare identification. Young stars are known to be more active than main sequence stars. Tracing the evolution of flaring activity can provide insight into the first few hundred million years of a star or planet's life.

\subsection{Selecting the Young Stellar Population}

The young stars selected for this sample are high-confidence ($> 50\%$) members of nearby young moving groups, young open clusters, OB associations, and star forming regions (see Table~\ref{tab:grouppaarameters}). Each of these stellar populations are comprised of coeval stars with common kinematics spanning a broad range of masses \citep{Zuckerman04}. The ages of the stellar populations evaluated in this study range from $\sim 1 - 800$\,Myr. Thus they are powerful tools for understanding stellar properties as a function of age and sample key early times in the formation and evolution of stars and planets.
    
After compiling a list of candidate young stars by aggregating selected \textit{TESS} Guest Investigator proposals and data from \citet{faherty18}, we used \texttt{astroquery} to access \Gaia\ DR2 to retrieve proper motions, parallaxes, radial velocities (when available) for the stars \citep{gaia16, gaia18}. We used derived effective temperatures, $T_{\rm eff}$, for the stars from the \tess\ Input Catalog (TIC) V8. We used \banyan\footnote{\url{https://github.com/jgagneastro/banyan_sigma}} to reconfirm and assign population memberships \citep{gagne18}. Using the stars' kinematics, \banyan\ derives a Bayesian probability of membership to 27 known coeval populations. We include stars with membership probabilities $\geq 50\%$, which yields a sample of 3193 stars observed at two-minute cadence observed in Sectors 1-20 (Table~\ref{tab:grouppaarameters}). In this sample, 2345 of the stars were observed in a single sector, 609 in two sectors, 103 in $\geq$ 3 sectors, and 7 in 13 sectors (the continuous viewing zone) of \tess\ data. 

\subsection{\tess\ Light Curve Pre-Processing}

\begin{figure}[!ht]
\begin{center}
\includegraphics[width=0.45\textwidth,trim={0.25cm 0 0 0}]{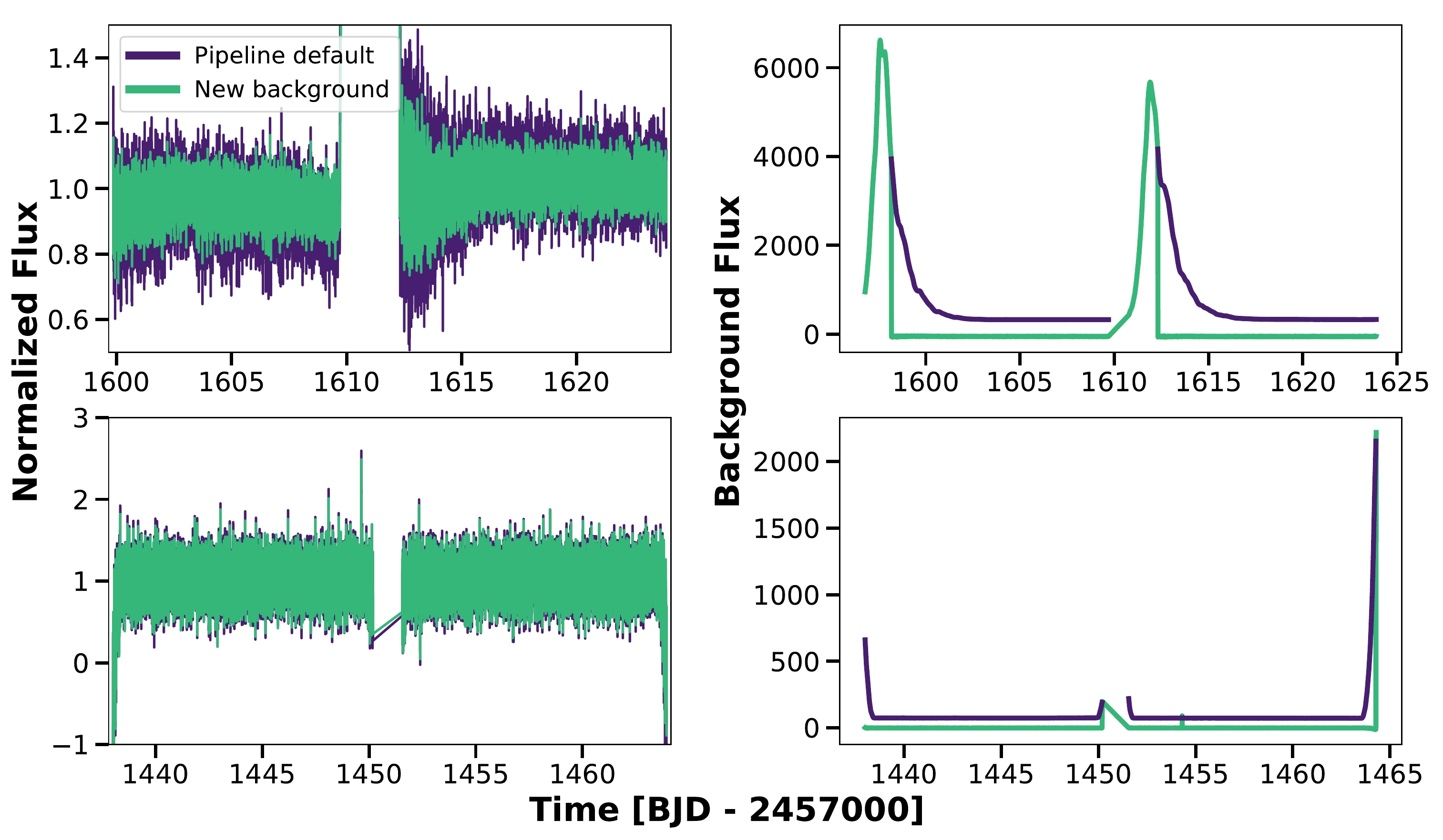}
\caption{Localized background subtraction \citep[green; based on][]{feinstein19} was performed to protect against background over-fitting (purple; default \tess\ pipeline background). Light curves are shown in the left column; background estimations are in the right column. The baseline background flux from the standard pipeline is greater than that of the localized background, suggesting over-fitting. Top is TIC 435801086 observed in Sector 11. Bottom is TIC 007652166 observed in Sector 5. \label{fig:diff_cdpp}}
\end{center}
\end{figure}

We used \lightkurve\footnote{\url{http://docs.lightkurve.org}} to download the target pixel files (TPFs) for all stars in our sample. In general, the Science Processing Operations Center (SPOC) pre-processed light curves were used with the pipeline default aperture. However, we found for several faint $T_{\rm mag}$ $> 14$  stars, the pipeline was returning higher than expected scatter or negative flux-value light curves. Specifically, we found the background model generated was over-correcting for some sources, seen in two examples in Figure~\ref{fig:diff_cdpp}, where the baseline background flux is not centered around 0. 

To mitigate this issue, and for consistency, we completed our own light curve background subtraction and aperture photometry. We follow the 1D background estimation in \cite{feinstein19}, subtract this background model from the 2-minute TPFs, and extract photometry using the SPOC identified aperture for all sources.

In addition, we use the pipeline-assigned quality flags to mask bad regions in the light curve. We use cadences with quality flags 0 and 512. Upon visual inspection, we found that quality flag 512 (``impulsive outlier removed before cotrending") at times removes the peak of flares across a range of amplitudes. To mitigate potential biases while calculating flare energies, we included this quality flag in the catalog. We additionally removed the first 200 cadences of each orbit to minimize Earthshine contamination.

\subsection{Measuring Rotation Periods} \label{subsec:rots}

\begin{figure*}[!ht]
\begin{center}
\includegraphics[width=0.8\textwidth,trim={0.25cm 0 0 0}]{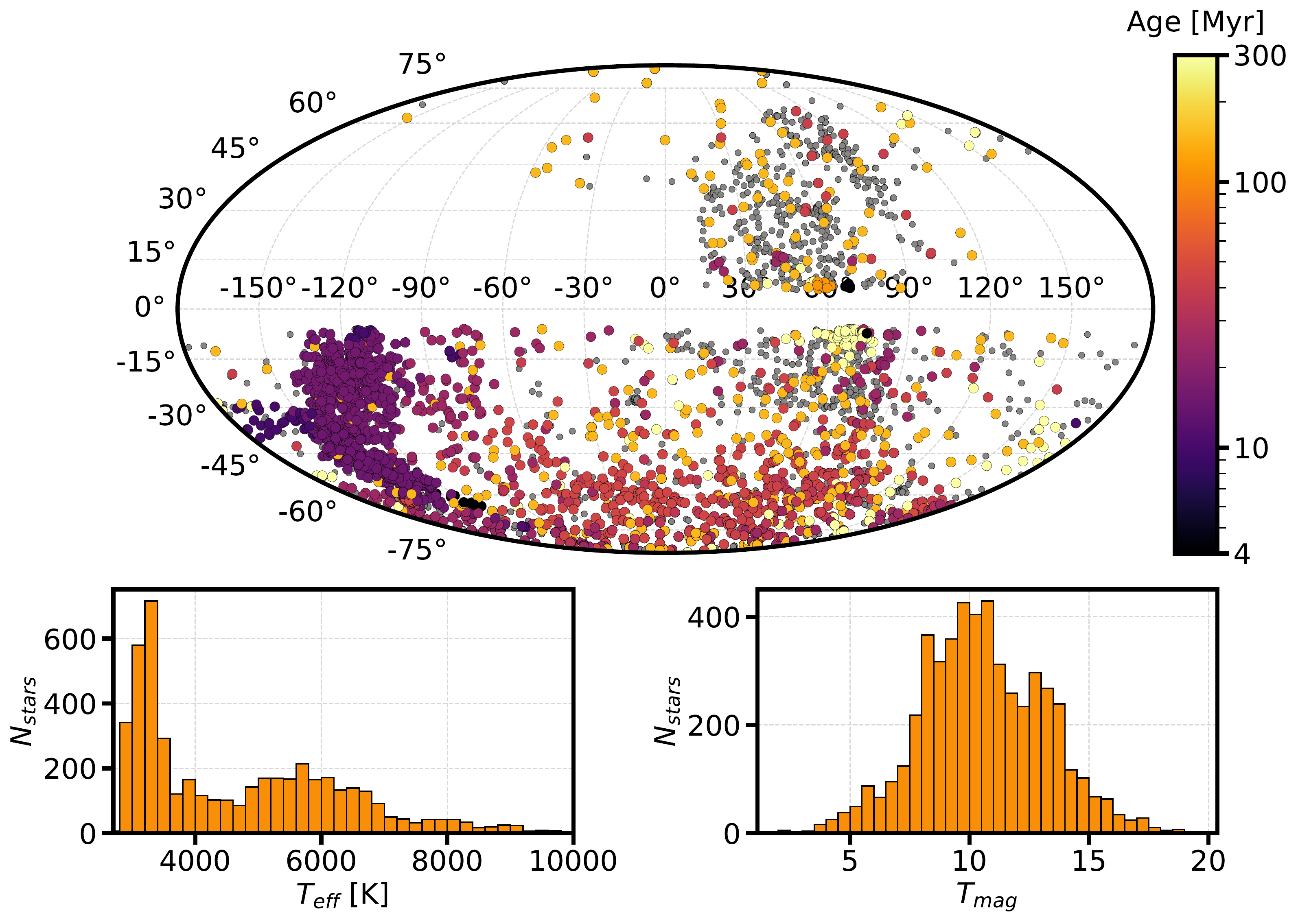}
\caption{The sample of young stars selected in this study. The top plot displays the locations, colored by age, of the sample across the sky in RA and Dec. Gray points which do not meet $> 50\%$ probability of membership to a given young population. The bottom left plot is the distribution of $T_{\rm eff}$. The bottom right plot is the distribution of \tess\ magnitudes, $T_{\rm mag}$. \label{fig:sample}}
\end{center}
\end{figure*}

One of the primary goals of this work is study the correlation of flare events and starspots. The rotation period ($P_{\rm rot}$) of a star can be measured through photometric spot variability. Thus, we created a sub-sample of stars with measurable rotation periods to explore the relationship between flares and spots. Rotation periods for stars observed in multiple sectors were identified per sector. We limited measured periods to $P_{\rm rot}< 12$ days; we found the contamination from Earthshine at the end of each orbit resulted in a strong periodic signal. Young stars are known to have short rotation periods, on the order of a few days, so this additional upper limit does not dramatically bias our sample \citep{Barnes03}. Statistics of the stars in our entire sample can be seen in Figure~\ref{fig:sample} and Table~\ref{tab:grouppaarameters}. 

\begin{figure}[!ht]
\begin{center}
\includegraphics[width=0.45\textwidth,trim={0.25cm 0 0 0}]{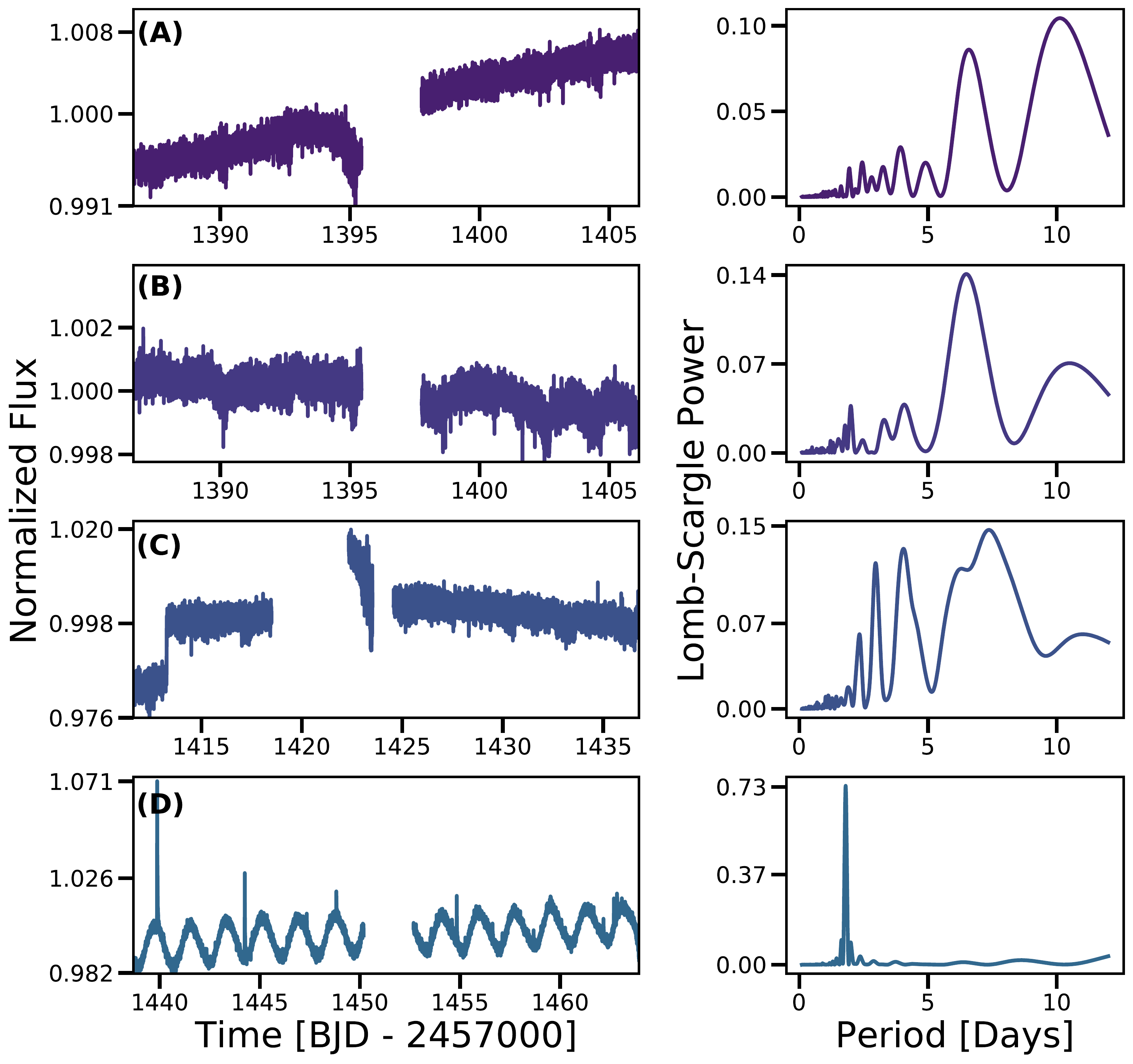}
\caption{Examples of light curves (left column) and corresponding Lomb-Scargle periodogram (right column) that did not pass one of the criteria used to find reliable periods. (A) TIC 3837491. There was no period measured for this light curve, thus resulting in a best-fit period of 12. We limit rotation periods to $P_{\rm rot} < 12$ days to avoid periodic Earthshine signals at the beginning or end of each orbit in a given sector (B) TIC 408017296. A period of $P_{\rm rot} < 12$ days was measured, however the width of the power peak is greater than expected, and there is no noticeable periodic variability in the light curve. This may be due to poor background correction and large Earthshine contamination. (C) TIC 250419751. The periodogram shows 2 potentially strong periods, even after masking resonances of the most likely rotation period. (D) TIC 1273249. A reliably measured rotation period with clear variability in the light curve. \label{fig:rot_cuts}}
\end{center}
\end{figure}

We used a Lomb-Scargle periodogram implemented in \texttt{astropy} to identify rotational signatures \citep{lomb76, scargle82}. We defined a set of criteria for the periodogram to pass in order to remove weak or non-existent rotation periods, highly unlikely rotation periods, or light curves that are dominated by noise or data gaps. The criteria as as follows: (A) the period must be less than 12 days; (B) the width of a Gaussian fit to the peak power must be less than 40\% of the peak period; (C) the secondary peak in the power spectrum must be 4\% weaker than the primary peak. Letters above correspond to a row in Figure~\ref{fig:rot_cuts} as examples for each criterion: each one reduces the possibility of catching a light curve that does not have a rotation period.

\begin{figure}[!ht]
\begin{center}
\includegraphics[width=0.45\textwidth,trim={0.25cm 0 0 0}]{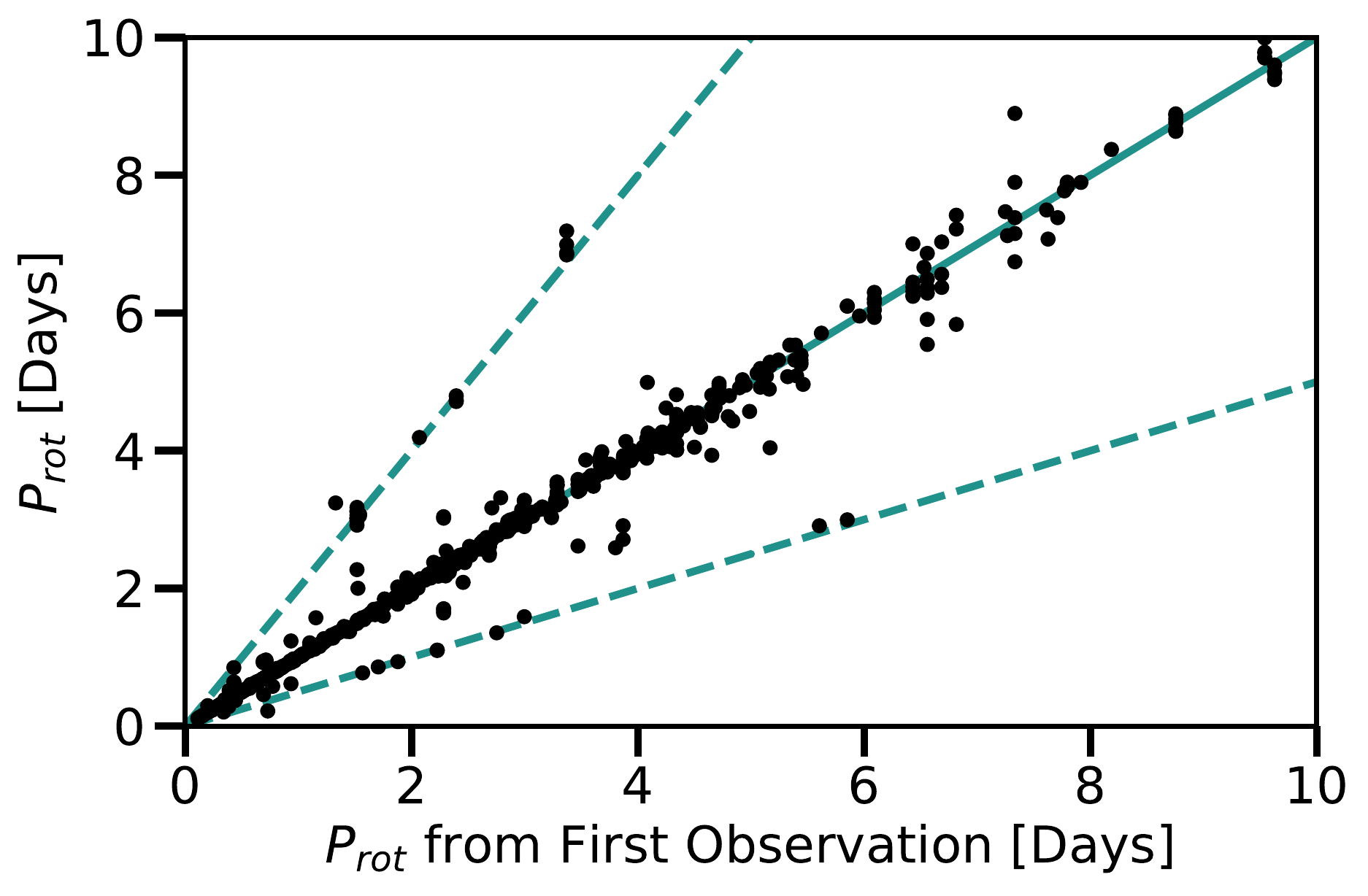}
\caption{A comparison of measured rotation periods of the same star observed across multiple sectors. A one-to-one guiding line is plotted in green. Vertically aligned points represent measured rotation periods from the same star. 85\% of rotation periods are consistent between sectors. Dashed lines represent periods that are twice and half the $P_{\rm rot}$ measured in the first observation. \label{fig:multi_sector_rotations}}
\end{center}
\end{figure}

Resonances of the peak period are masked before identifying the second strongest period in the periodogram. For stars observed in multiple sectors, we compare the rotation periods measured from each sector to the mode from all sectors. Additionally, we searched the second strongest power in the periodogram as we found, on occasion, the strongest period was caught by contamination in the light curve. Half the period and twice the period were each compared to the mode of the sample; if the period agreed within 0.1 days, it was considered a reliable measurement. In total, we measured rotation periods for 1513 stars. The resulting rotation periods for stars observed in multiple sectors can be see in Figure~\ref{fig:multi_sector_rotations}. The rotation periods are in good agreement across different sectors, sometimes varying as half or twice the periods, which is accounted for in the final median rotation period. 

\begin{figure}[!ht]
\begin{center}
\includegraphics[width=0.45\textwidth,trim={0.25cm 0 0 0}]{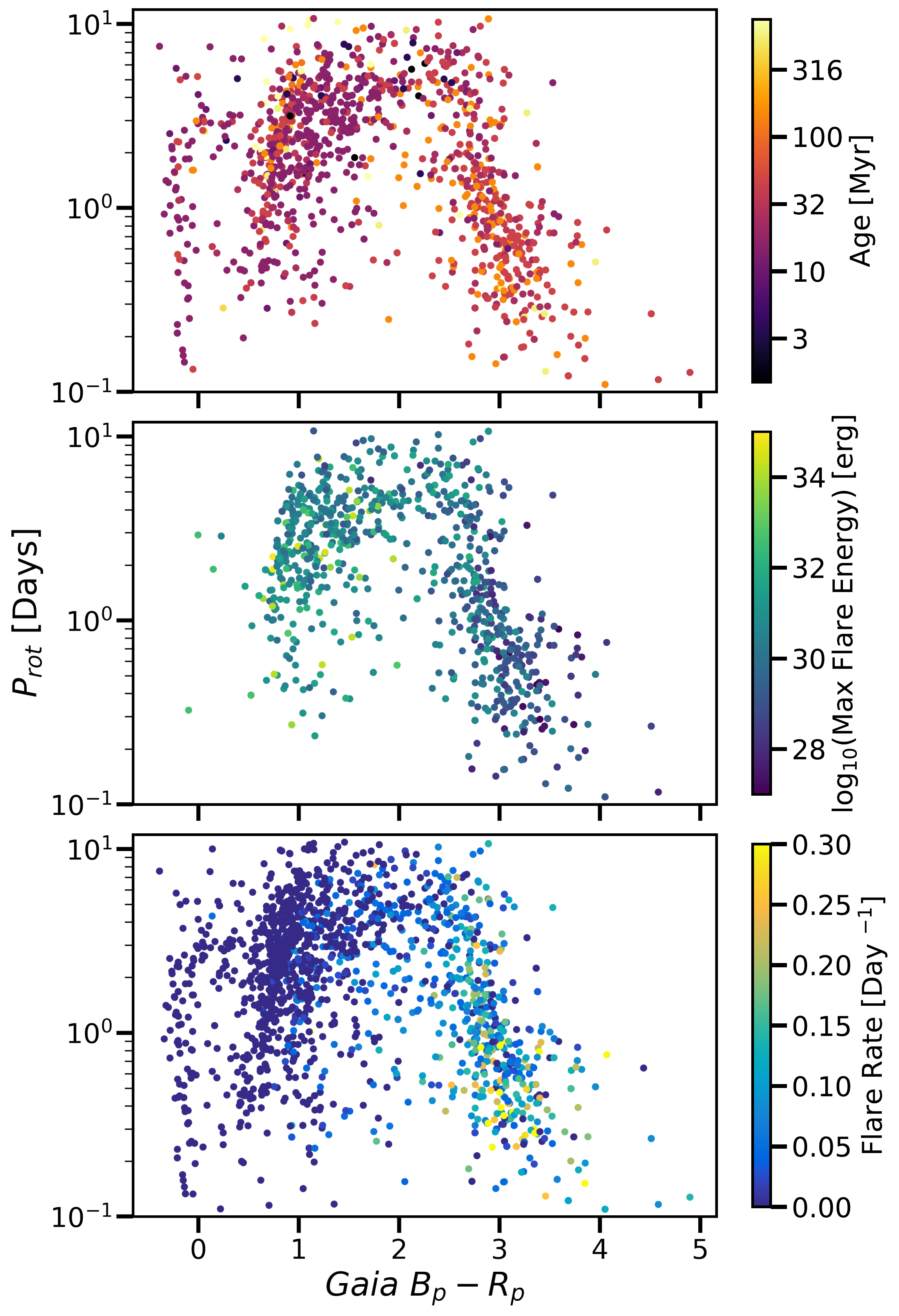}
\caption{The rotation period distribution as a function of \Gaia\ $B_p-R_p$ color. The top plot is colored by age; the middle plot is colored by maximum flare amplitude found on that star; and the bottom plot is colored by flare rate, where the flares are weighted by probability assigned by the CNN. $B_p-R_p$ = 2 roughly corresponds to $T_{\rm eff} \approx$ 4000K, as provided by \Gaia.  There is a clear drop-off in large flares at $T_{\rm eff} \approx$ 4000K (bottom), where hotter stars do not have as strong flares, regardless of $P_{\rm rot}$. \label{fig:prots}}
\end{center}
\end{figure}

The rotation periods for all stars are plotted as a function of \Gaia\ $B_p-R_p$ color and colored by age in Figure~\ref{fig:prots} (top). There is a clear artificially induced break at $P_{\rm rot} > 12$ days, as a result of our rotation period metrics. Stars $> 9000$K were visually inspected for signs of rotation periods. It is possible that spot variability are from a binary companion to these hot stars, however for this analysis we assume this is not the case. The elbow at $B_p-R_p$ = 2 is seen in other gyrochronology studies \citep[e.g.][]{curtis19}. The most likely associated young population membership were assigned by \banyan\ and ages for each population is listed in Table~\ref{tab:grouppaarameters}.

\subsection{Identifying Flares}

\begin{figure*}[!ht]
\begin{center}
\includegraphics[width=\textwidth,trim={0.25cm 0 0 0}]{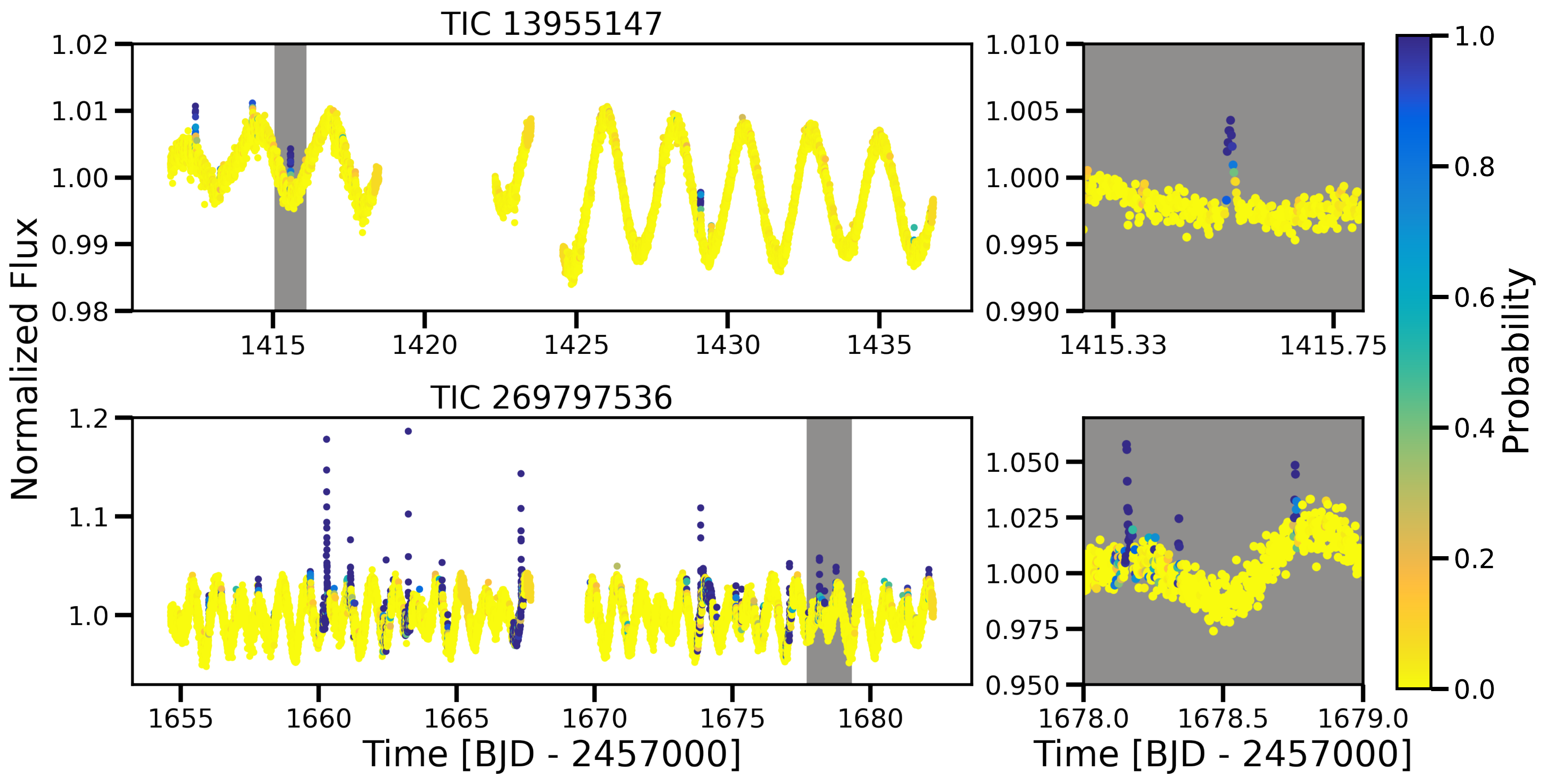}
\caption{Examples of light curves colored by the average ``probability" as determined by the ten ensembeled CNN models. An individual flare from each star is highlighted and displayed in the right column. The probability the cadence is part of a flare increases from yellow to purple. Orange areas at the beginnings and ends of continuous observations were ignored by the CNN due to large gaps in the data. Flares are seen to be easily distinguishable from differing spot modulation, which consistently has a low flare probability. \label{fig:example_lc}}
\end{center}
\end{figure*}

The data used to train and validate the CNN required the flare to be at the center of each example. To predict where flares occur in other light curves, we used the CNN as a sliding box (with a baseline of 200 cadences) detector where each cadence is centered within the new examples, similar to the methods of \cite{pearson18}. This removed biases for finding flares that may or may not have occurred in the center of an example if we evenly divided the light curve into 200 cadences. As flares appear towards the center of the sliding box, the probability assigned to that cadence increases, as seen in Figure~\ref{fig:example_lc}. Each light curve was fed into the 10 \stella\ models described in Section~\ref{subsec:cnn-results}. The output predictions from each model were averaged per light curve and used for flare identification. 

We considered every point in the ``probability"-light curve above a threshold of 0.5 to be a potential flare. Consecutive cadences of greater than the threshold were considered to be part of the same flare. The amplitude of the flare is defined as the point in the series with either the highest probability or, if several cadences share the same probability, the corresponding maximum normalized flux. Examples of identified flares can be seen in Figure~\ref{fig:example_lc}, where cadences are colored by the averaged probability identified with our ten CNN models. Both light curves were fed into the CNN as is. As the CNN requires 200 sequential cadences, without data gaps, the first and last 200 cadences of each orbit are ignored (Figure~\ref{fig:example_lc}, ``large" regions orange points).

Per each flare, we extract parameters by computing a $\chi^2$-fit to an empirical flare model \citep{walkowicz11, davenport14}. We chose to model the flares by a sharp Gaussian rise and an exponential decay. The localized underlying stellar variability was removed before fitting the flare model by linearly interpolating across 100 cadences before and after the flare, where the flare itself is masked, and dividing this model from the light curve. Although flares sometimes show more complex structure, we found this simple model to result in reasonable fits to the flare amplitudes.

The CNN provides a good initial guess for where flares are located. However, due to a lack of completeness for flares down to low energies in the training set, we include minor additional checks that the event was a potential flare. Such checks include: (1) the amplitude of the flare must be $1.5\sigma \geq$ the locally detrended light curve, (2) the two cadences directly following the amplitude must be $1\sigma \geq$ the detrended light curve (i.e. at least 3 consecutive outlier points are considered part of a flare), (3) the cadence before and after the flare amplitude must be less than the amplitude, and (4) the amplitude must be $\geq 0.5\%$.

\section{Results}\label{sec:results}

In total, we identified 17,479 flares across the 3193 stars in our sample. Of these, 14,618 had amplitudes $< 5\%$ in the \tess\ bandpass. Figure~\ref{fig:prots} shows \Gaia\ color plotted against $P_{\rm rot}$ colored by age (top), maximum flare energy (middle), and flare rate (bottom). There is a noticeable decrease in flare rate at \Gaia\ $B_p-R_p < 2$, where  $B_p-R_p = 2$ corresponds to $T_{\rm eff} \approx 4000$K. For Sun-like stars and hotter, there is a clear drop-off in flare energies, with many stars not experiencing any flares and are thus absent from the middle plot. \cite{notsu19} found a decrease in maximum flare energy with increasing rotation period for \kep\ stars. Additionally, \cite{tu20} observed the maximum flare energy as a function of rotation period for Solar-like stars in \tess\ data and found similar results. Magnetic activity is heightened for shorter rotation periods; it is interesting to note that we find no such relationship here with regards to the maximum flare energy as a function of rotation period. All young stars of the same color show similar flare energies across rotation periods.

\subsection{Flare Rates as a Function of Age and Temperature}

\begin{figure*}[!ht]
\begin{center}
\includegraphics[width=\textwidth,trim={0.25cm 0 0 0}]{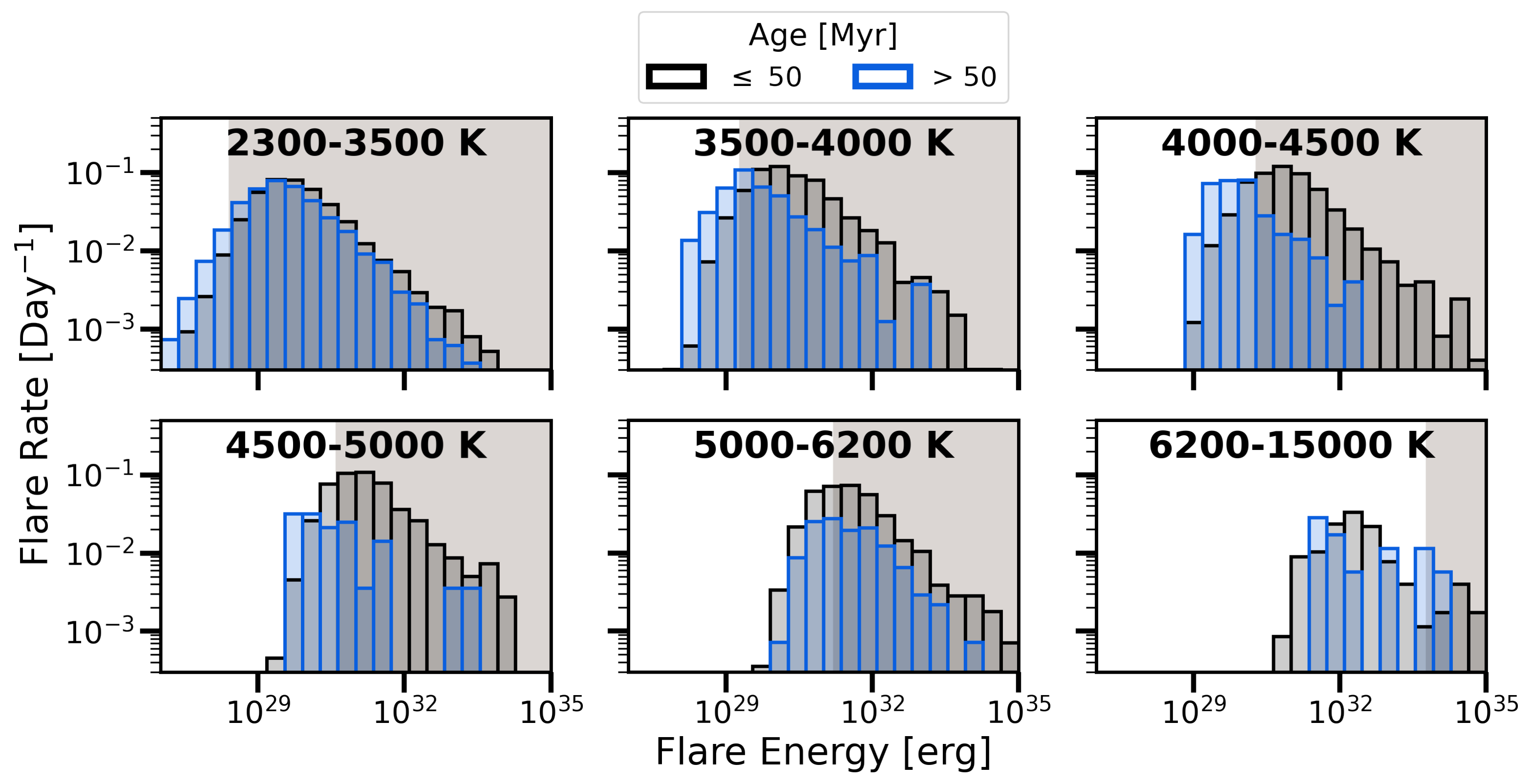}
\caption{Flare rates broken down by effective temperature and colored by age, where black bins represent stars $t_{\rm age} \leq 50$\,Myr and blue bins represent $t_{\rm age} > 50$ Myr. The gray shaded region corresponds to the energies at which we expect to be able to detect flares on all stars in that sub-panel. The flares are weighted by their assigned probability from the CNN. Temperature ranges are labeled in each subplot. There is a noticeable drop-off in flare rate and energy as the star's temperature increases. M and late K dwarfs ($2300 \leq T_{\rm eff} \leq 4000$ K) experience similar flare rates and energy across the entire range of our sample. We do not recover any flares on stars $6200 \leq T_{\rm eff} \leq 15000$ K older than 50 Myr.} \label{fig:agedist}
\end{center}
\end{figure*}

The flare rates are evaluated as a function of both stellar effective temperature and age. Figure~\ref{fig:agedist} is binned by flare energy and colored by ages $t_{\rm age} \leq 50$ Myr (black) and $t_{\rm age} > 50$ Myr (blue); each subplot represents different spectral types. All flare analysis was weighted by the probability assigned by the CNN, as a flare with a 50\% probability has an equal chance of being a flare or a non-flare. This figure is additionally binned in log-space and range in energies from $10^{29}-10^{35}$\,erg. The coolest stars exhibit greater flare rates across all energies and many more low-energy flares, with a tail extending out to $10^{35}$\,ergs. There is a detection bias against low-energy flares with increasing $T_{\rm eff}$; the amplitudes of low-energy flares are smaller on hotter stars and therefore are undetectable. Our minimum flare amplitude is 0.5\%, which has an equivalent duration (ED), or area under the flare light curve, $\sim 1.5$\,s. For each temperature bin, we calculated the corresponding energy to such a flare. We found a relation of $E_{\rm flare, min} = 0.01\, T_{\rm eff}^{8.65 \pm 1.61}$.

The tail extends across all ages, ref{fig}bug with a more dominant presence in the $\leq 50$\,Myr stars. No flares were seen on $T_{\rm eff} \geq 6200$\,K and age $t_{\rm age} >$\,50\,Myr. $T_{\rm eff} \sim$ 6200\,K roughly corresponds to the Kraft break \citep{heger00}. Thus a change in internal structure may be the cause of such a lack of flares, although we note our sample has very few stars in this bin. There are only 132 stars in this temperature range, and only 3 of these are older than 50 Myr.

\begin{figure*}[!ht]
\begin{center}
\includegraphics[width=\textwidth,trim={0.25cm 0 0 0}]{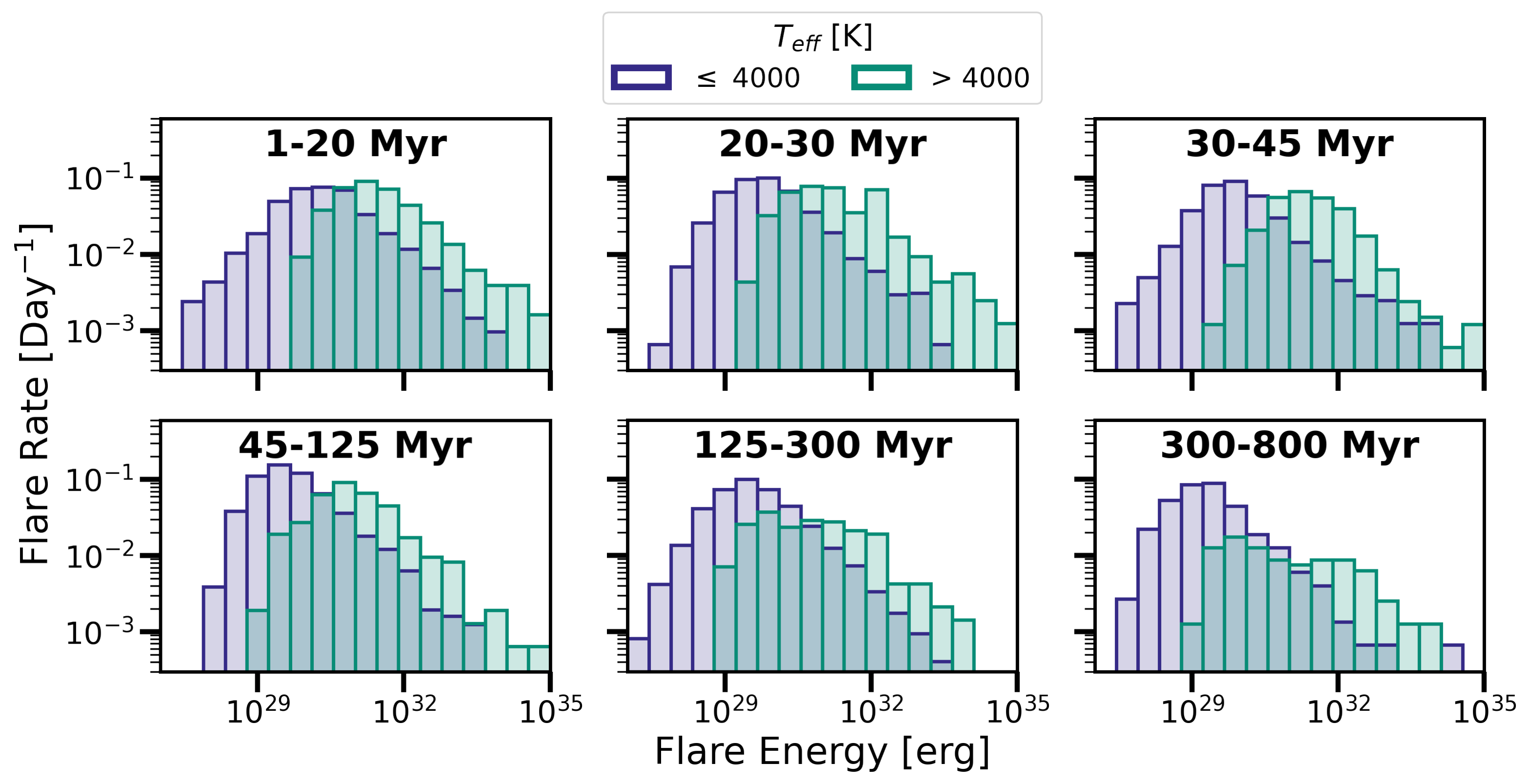}
\caption{Flare rates for our sample broken down by age and colored by effective temperature, where purple bins represent $T_{\rm eff} \leq 4000$K and green bins represent $T_{\rm eff} > 4000$K. The flares are weighted by their assigned probability from the CNN. Cooler stars exhibit more low-energy flares due to detection biases. However, across all sub-panels, the cool stars show greater than or equal to flare rates compared to the hotter stars.} \label{fig:teffdist}
\end{center}
\end{figure*}

Figure~\ref{fig:teffdist} evaluates flare rates by temperature, that which matches the maximum flare rate drop-off in Figure~\ref{fig:prots} (bottom panel), and binned in age ranges. Across all age ranges, stars hotter than 4000\,K (green) uniformly display smaller flare rates. The cooler stars have relatively consistent flare rates for both low and high-energy flares in each age bin, demonstrating an overall level of heightened activity for these stars. There may be a slight increase in high-energy flares and flare rates between 1-20\,Myr and 30-45\,Myr stars cooler than 4000\,K. This may be due to the fact that low-mass stars evolve more slowly onto the main sequence. These results are similar to those presented in \cite{ilin19}. However, the ages explored in this sample are much younger than theirs.

Additionally, we combined the subplots of Figure~\ref{fig:teffdist} into flare frequency distributions (FFD) for $T_{\rm eff} \leq 4000$\,K and $T_{\rm eff} > 4000$\,K. A power-law was fit to each FFD, with the same bin sizes. The slope of each was found to be $\alpha_{\leq 4000} = -1.59 \pm 0.03$ and $\alpha_{> 4000} = -1.57 \pm 0.04$, which are within a $1.5\sigma$ agreement between the samples. We repeated the same calculation, but instead by fitting the FFD for $t_{\rm age} \leq 50$\,Myr and $t_{\rm age} > 50$\,Myr. We find the slope of each to be $\alpha_{\leq 50 \rm Myr} = -1.58 \pm 0.03$ and $\alpha_{> 50 \rm Myr} = -1.52 \pm 0.08$. The FFDs binned by age are likely dominated by the flare rate of the coolest stars. With \tess's extended mission, more data on same and different young stars may reveal a more significant difference in FFD slope as a function of temperature and age.

\subsection{Correlation between Spots and Flares}

The phase of the light curve was then mapped using the best-fit or averaged (if observed in multiple sectors) rotation period. One full rotation period has a range of phases: $-0.5 < \phi < 0.5$, where the peak of the spot modulation is at $\phi = 0$. The phase was mapped at each orbit. Partial rotations near the beginning/end or orbit gap were extrapolated from the surrounding full rotation periods. 

\begin{figure}[!ht]
\begin{center}
\includegraphics[width=0.45\textwidth,trim={0.25cm 0 0 0}]{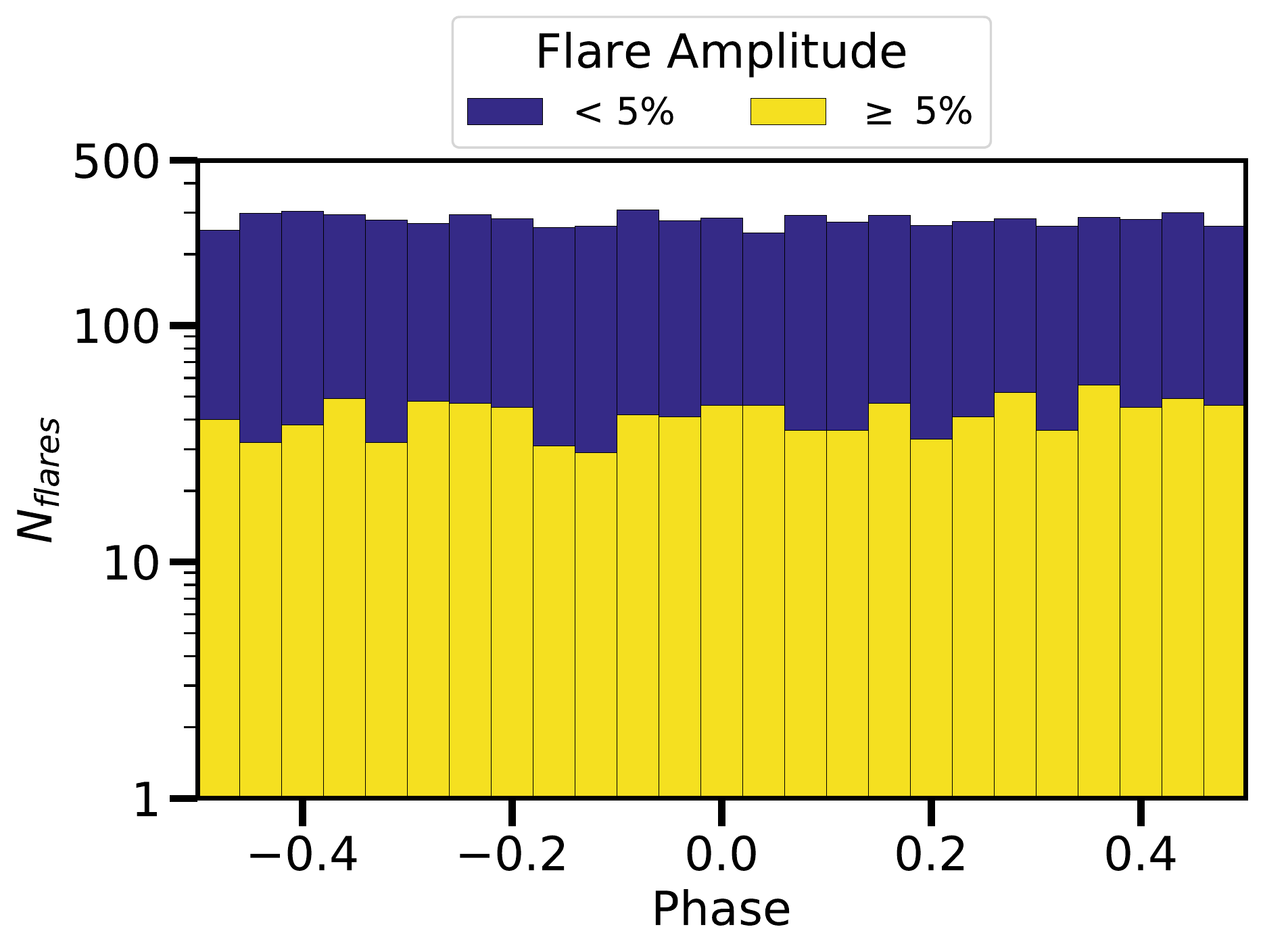}
\caption{Histogram of flares as a function of phase along a light curve. Yellow are super-flares ($\geq 5\%$ flux increase) and purple are $< 5 \%$ flux increase. Flares are seen across all phases, which suggests both hemispheres could host spot groupings and the variability seen is the differences in spottiness. Bin sizes correspond to $\phi = 0.04$. \label{fig:flare_phase}}
\end{center}
\end{figure}

Figure~\ref{fig:flare_phase} shows a histogram of flares with respect to phase, where the $\phi$ bin sizes are $0.04$. Low- ($<5\%$, purple) and high- ($\geq 5\%$, yellow) amplitude flares are separated. For both samples, there is a consistent spread in number of flares at each phase. These results are broadly consistent with those of \cite{notsu13, doyle18, doyle19, doyle20}. These previous studies have looked at flare rates versus phase for samples of size $\sim 100-200$, here we find the same effect with sample size 1500.  Taken together, we interpret Figure~\ref{fig:flare_phase} as evidence for magnetic active region coverage that is large and mostly uniformly distributed in longitude.

Starspots are dark regions on a star originating from a concentration of magnetic field lines in the photosphere. As such, it is believed that flares will likely occur near these concentrations, as seen on the Sun \citep{dun07, zhang08}. By fitting a sine-wave to the overall sample of flares presented in Figure~\ref{fig:flare_phase} at the same phase as the x-axis, we find a 3\% difference in the flares between $\phi = 0$ and $\phi = \pm 0.5$. This small phase dependence means that the projected filling factor of stellar active regions remains relatively constant as seen from our viewing location.  The on-average 2\% spot-induced variability seen in the \tess\ light curves must be the result of a stars covered in starspots, with only slight differences between the most-spotted and least-spotted projected hemispheres. 

In the \tess\ bandpass (600-1000\,nm), the Sun would be observed to have part-per-thousand variability. The surfaces of the stars in our sample would be very different compared to the Sun. Stellar active regions on this sample of young stars exhibit a large coverage fraction, with a large degree of longitudinal symmetry.  Such symmetries can arise from potentially few, but large spots, very many small spots, circumpolar or polar spots on inclined stars, or active latitudes peppered with spots \citep{rackham18, guo18}. The latitudinal distribution of starspots as a function of age is generally unknown, with evidence for circumpolar starspots on both young and evolved stars \citep{2009ARA&A..47..333D, 2016Natur.533..217R}.  Such high-latitude structures may act to mask the mapping from lightcurve amplitude to total starspot coverage \citep{guo18} as spots remain in view on moderately inclined stars.  Such viewing angle effects could explain the relatively uniform frequency of flares with rotational phase. 

GJ 1243 is a known older active M dwarf and was found to have no correlation between starspots and flares \citep{Hawley14}. That paper proposed two explanations: that there is either significant spot coverage or a large polar spot consistently in the field of view. While polar spots would have different effects as a function of inclination, the effects of high spot coverage would look broadly similar across a large sample of stars. As our sample contains a large selection of stars likely at a variety of inclinations, and we see no significant correlations across our sample, our results favor the high spot coverage scenario for GJ 1243 and other active systems.

Nearly 10 years of observations of the T Tauri star V410 Tau concluded in a relative coverage from 29-41\% coverage on the less spotted hemisphere to 61-67\% on the more spotted hemisphere \citep{grankin99}. LkCa 4 was found to have a total starspot coverage fraction of 74-86\% on LkCa~4, for an asymmetric-to-total spot coverage ratio of 15\% \citep{gully17}. Detailed studies of low-mass members of the Pleiades ($\sim 112$\,Myr) suggest $\sim 30\%$ and $\sim 50\%$ spot coverage may be common for young M and K dwarfs, respectively \citep{stauffer03, fang16}. With such varying differences of spot coverage star to star, but with consistency in significant coverage up to 112 Myr, detailed studies of each star with spot modulation would need to be completed to truly understand the underlying distribution of spots.

However, understanding the coverage fraction of spots does not provide insight into the shape and nature of the spots. \cite{rackham18} modeled light curves with varying spot coverage levels with both small (Sun-like, $R_{spot} = 2^\circ$) and large ($R_{spot} = 7^\circ$). For the similar level of spot coverage, the large spots resulted in higher spot variability amplitudes than that of the small spots. They also concluded the relationship between spot coverage and observed variability is nonlinear.  

\cite{morris2020} estimated the spot coverage of 531 F-K type stars across 10 Myr - 4 Gyr finding spot coverage decreases exponentially as $\propto t^{-0.37}$, where $t$ is stellar age given in Gyr. For the sample presented here, this would suggest a spot coverage of $1-13\%$ for the oldest to youngest stars. Even when broken down by age, Figure~\ref{fig:flare_phase} remains relatively constant across phase. This suggests that even at older ages, the total spot coverage fraction must be very large and may be underestimated from light curve-based methods, which are largely only sensitive to hemispherical asymmetries and rely heavily on uncertain geometrical assumptions on the longitudinal spot distribution.

\section{Discussion}\label{sec:discussion}

\subsection{Comparing to Previous Flare Identification Methods}

We completed an injection-recovery test to compare the results of the CNN to previous methods. \stella\ was trained on real flare events;  injected flares do not perfectly model such events. Therefore, we do not expect the performance of \stella\ on this sample to be identical to the performance on real data, and the results of this test should be viewed as an imperfect estimate of the effectiveness of the predictive model. The results of this test thus do not provide a definite result of how well \stella\ is doing compared to previous methods. In order to completely quantify the differences, a more detailed study would need to be completed, which is beyond the scope of this analysis.

The injected flare amplitudes were randomly chosen from a normal distribution centered around a 4\% flux increase from the baseline light curve. Ten flares were injected in 2263 of the light curves, those with measured rotation periods, in our sample. The injected light curves were then processed using the \stella\ CNN and flare identification module. To compare to previous methods, we chose to follow the methods of \cite{chang15} (Equations 3a-d), which have been used for flare analyses for both \kep\ and \ktwo\ data \citep{Davenport16, ekaterina19}. We used a Savitsky-Golay filter of window length = 15 \citep{Davenport16}, as implemented in \lightkurve, to detrend the spot modulation in the light curves before applying the flare heuristics. An aggressively small kernel size was used to combat the range of rotation periods measured in the sample.

We used a probability threshold of 0.5 for positively identified flares with \stella. The injected flares may not perfectly represent the real astrophysics behind a flare, which the CNN was originally trained on. As such, an injection recovery provides useful guidance for comparison between methods, and should \textit{not} be interpreted as a definitive efficacy. Nevertheless, the injected flares can provide some insight into overall trends in flare detection. While both methods recover all flares of amplitudes greater than 11\% flux increase, there is a steep drop-off in recovered low-amplitude and generally lower energy flares for previous methods. The CNN is able to recover 80\% of injected flares below 5\% flux increase. The resulting distribution of recovered flares with \stella\ is similar to the injected sample, while previous methods are more centered at higher-flare amplitudes and do not accurately represent the underlying distribution of injected flares. 

\subsection{Largest Discovered Flares}

\begin{figure}[!ht]
\begin{center}
\includegraphics[width=0.45\textwidth,trim={0.25cm 0 0 0}]{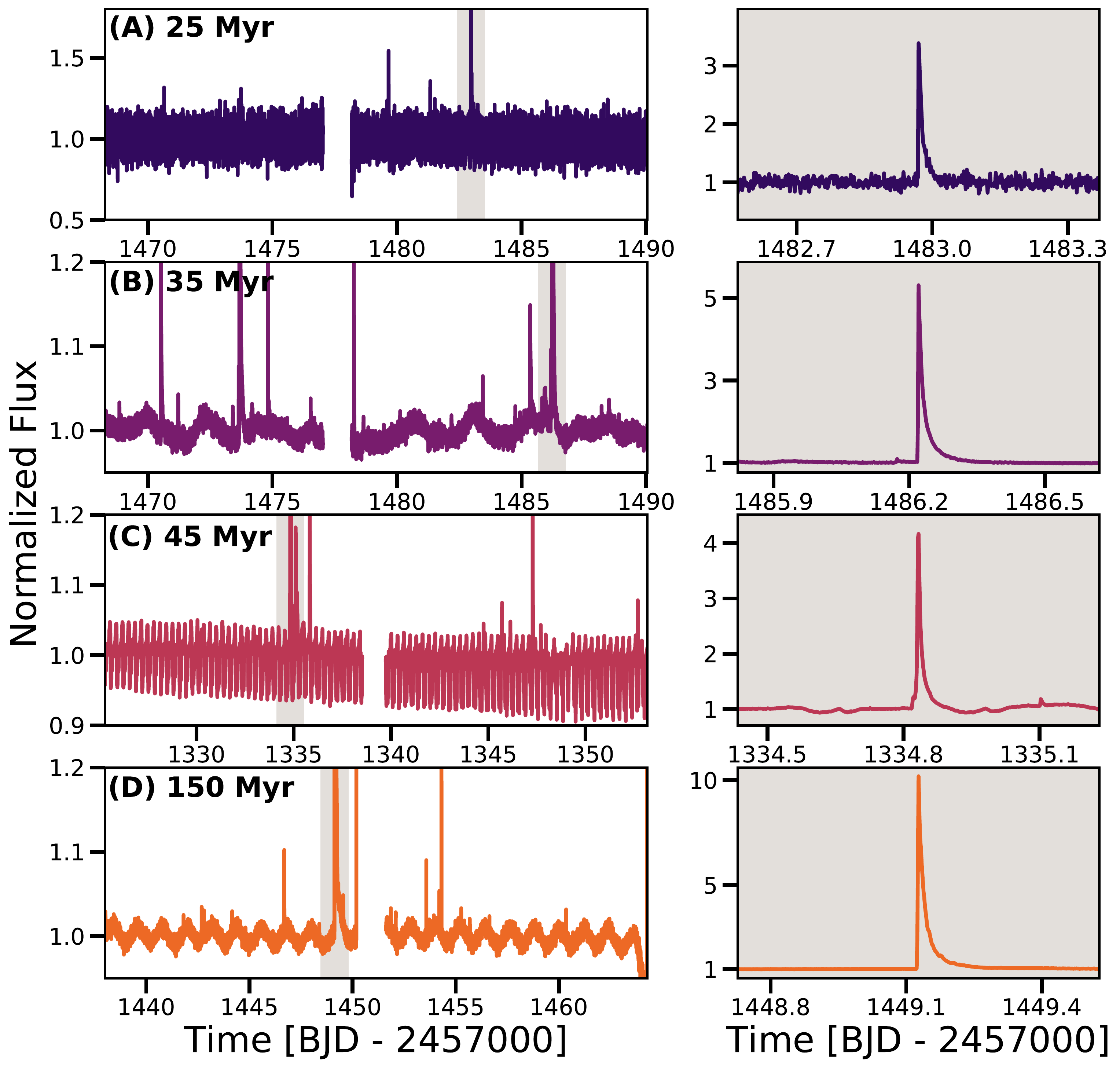}
\caption{A few examples of the largest recovered flares in the sample in increasing age order. The flares are highlighted in gray and plotted with a different y-scale in the right column. (A) TIC 24721262 is a member of the $\beta$ Pictoris moving group with $T_{\rm eff} = 2836$K and $L/L_\odot = $ 0.001. (B) TIC 77954300 is a member of the Octans association with $T_{\rm eff} = 3171$K and $L/L_\odot = $ 0.017. (C) TIC 206544316 is a member of the Tucana-Horologium association with $T_{\rm eff} =  3237$K and $L/L_\odot = $ 0.021. (D) TIC 44678751 is a member of the AB Doradus moving group with $T_{\rm eff} =  3201$K and $L/L_\odot = $ 0.011. Note the changes in y-axis, with the largest flare occurring on the oldest star of this sub-sample. Stellar parameters were obtained from TIC V8. \label{fig:large_bois}}
\end{center}
\end{figure}

A selected sample of some of the highest amplitude flares recovered in this work are shown in Figure~\ref{fig:large_bois}. The light curves have been normalized and are labeled by the best age estimation for that star. The best-fit young population membership and stellar parameters are presented in the figure caption. There is clear underlying spot modulation for most of the stars plotted (left column) and the superflare is highlighted with a different y-scale limit to the right. These four stars span a temperature range of $T_{\rm eff} = 2800-3200$ K and luminosity values between $0.001-0.02 L_\odot$, as provided in the TIC and are representative of all stars with flares amplitudes $\geq 100\%$. 

The ED and the luminosity can be combined to estimate flare energy in the \tess\ band. We use the simple flare model described previously to estimate the ED. Note that the flare models used are a rough estimate. Large flares are sometimes found to be followed by lower-energy sympathetic flares, which extend the exponential decay \citep{davenport14}. The models used to estimate the ED and energy of these flares does not include this potential feature. 

The resulting flare energies, in ascending age order for the targets displayed in Figure~\ref{fig:large_bois} (TIC 24721262, TIC 77954300, TIC 206544316, and TIC 44678751), are: $3\times10^{35}$, $1\times10^{37}$, $7\times10^{36},$ and $2\times10^{37}$ erg. The `Carrington' superflare event on the Sun in 1859 released $\sim 10^{32}$ erg, which is several orders of magnitude weaker than the strongest flares found in this work \citep{carrington1859, hodgson1859}. Using the V, J, H magnitude relation in \cite{Stassun17} and the measured flare amplitude, we find the change in magnitude for these stars during the flares to be $\Delta V =$ 3.22, 4.11, 3.52, and 4.21 for the latter three targets. Flares are treated as a 9000\,K blackbody \citep{kretzschmar11}. As there was no V magnitude available for TIC 24721262, we used the G magnitude as an estimation.

\subsection{Repercussions for Exoplanets}

As the confirmed number of low-mass stars hosting exoplanets increases \citep[e.g.][]{feinstein19_planet, gillon17, guenther19_exoplanet, luger17, Kostov19}, so too does the need to understand the environment in which these planets grow and reside. High flare rates and high-energy radiation of main sequence stars have raised questions about the habitability of these planets. It could inhibit the evolution of life through the suppression or destruction of an atmosphere \citep{tilley19}. Or, flares could create prebiotic molecules to ignite complex chemistry required for life and allow for increased greenhouse warming \citep{airapetian16}.

Both the frequency and energy of the flares can affect exoplanet atmospheres. Simulations of the flares from AD Leo, an active M dwarf, on two different modeled planets found that a single high-energy flare \citep[E$\sim$10\textsuperscript{34}erg;][]{hawley85} was sufficient enough to irreversibly alter the chemical composition of the atmospheres \citep{venot16}. Specifically, the relative abundances of hydrogen, ammonia, carbon dioxide, nitric oxide, and hydroxide as a function of atmospheric pressure were significantly different $\sim30,000$ years after a flare was injected, concluding that planets around very active stars are constantly and permanently altered by these events. The highest-amplitude flares in Figure~\ref{fig:large_bois} are 1-3 orders of magnitude larger than the flare for AD Leo. Although \cite{segura10} found the enhanced UV radiation did not affect the habitability of an Earth-like planet around AD Leo, they did not account for short-duration high-energy flares. \cite{vida17} analyzed the flares produced by TRAPPIST-1 in the available \ktwo\ light curve and concluded that both the high frequency and energies are likely disadvantageous for life.

Only a handful of transiting exoplanets around young stars are known \citep[e.g. K2-25, K2-33, DS Tuc Ab, and V1298 Tau bcde;][]{mann16, david16, benatti19, david19_v1298all, david19_v1298b, newton19}. The atmospheres of these planets are believed to be highly extended and rapidly evolving. The ``radius-gap" in \kep\ planets at $\sim 1.6 R_\oplus$ has been theorized to the be result of significant photoevaporation of atmospheres within the first 100 Myr \citep{owen17}. The models from \citep{owen17} considered a higher background UV environment for the exoplanet atmospheres, but did not include flares. The results presented here suggest high flare rates and amplitudes within the first 125\,Myr for stars $T_{\rm eff} > 4000$\,K. Cooler stars demonstrate high rates across 800\,Myr. 

We combined the methods described in \cite{owen17} and \cite{owen20} to study the evolution of atmospheric mass of an exoplanet over time. Additionally, the equations were modified to account for ``flare-like" events, or short bursts of high-energy luminosity. We evaluated the potential impact of flares on V1298\,Tau\,d, a 23\,Myr old $6.41 R_\oplus$ planet at $a = 0.108$\,AU separation discovered with \ktwo\ \citep{david19_v1298all}. The following additional assumptions about the planet were made: (1) the iron fraction in the core, $X_{iron} = 1/3$; (2) the ice fraction in the core, $X_{ice} = 0$; (3) the mass of the core, $M_{core} = 5 M_\oplus$; (4) the mass loss efficiency, $\eta = 0.2$. The mass-loss rate due to photoevaporation is then given by

\begin{equation}
    \dot{M} = \eta \frac{\pi R_p^3 L_{HE}}{4 \pi a^2 G M_{core}} f(A),
\end{equation}

\noindent where $L_{HE}$ is the luminosity of high-energy photons from the star, $R_p$ is the radius of the planet, $a$ is the semi-major axis \citep{owen17}, and $f(A)$ is a factor of mass-loss rate increase as a function of flare amplitude, $A$. This factor in the presence of a flare follows the relations found in \cite{bisikalo18}, and takes the form of $f(A) = 0.95A^{2.6}$. We assumed a $L_{HE}$ following Equation 22 in \cite{owen17}, which remains constant for the first 100\,Myr and decays afterwards as $\rm t_{age}^{-1.5}$.  Although several studies have shown a correlation between flare rate and $L_{HE}$ \citep{maehara17, notsu19}, these were for solar-type stars and may not apply to young K dwarfs like V1298\,Tau. Flare amplitudes were injected up to 100$\times$ the base luminosity as observed in the \tess\ bandpass. We include an additional parameter, $s$, which converts the observed flare amplitudes in \tess\ to the amplitude observed in broad UV wavelength range and was calculated by

\begin{equation}
    s = \int_{TESS} \frac{B_\lambda(T_{\rm eff})}{B_\lambda(T_{\rm flare})} d\lambda \left[\int_{UV} \frac{B_\lambda(T_{\rm eff})}{B_\lambda(T_{\rm flare})} d\lambda\right]^{-1}
\end{equation}

\noindent where $B_\lambda(T)$ is the Planck function, the \tess\ bandpass covers $600-1000$\,nm, and the UV covers $66-365$\,nm. We have assumed all flares can be represented as a 9000K blackbody \citep{kretzschmar11}. For V1298\,Tau, where $T_{\rm eff} = 4970$\,K \citep{david19_v1298b}, $s = 14.6$, which indicates flares would appear $14.6\times$ brighter if observed in the UV. At each time step, there was a $\sim 33\%$ chance of a flare being injected, which is a similar probability to observing a flare in the \tess\ light curves.

\begin{figure}[!ht]
\begin{center}
\includegraphics[width=0.45\textwidth,trim={0.25cm 0 0 0}]{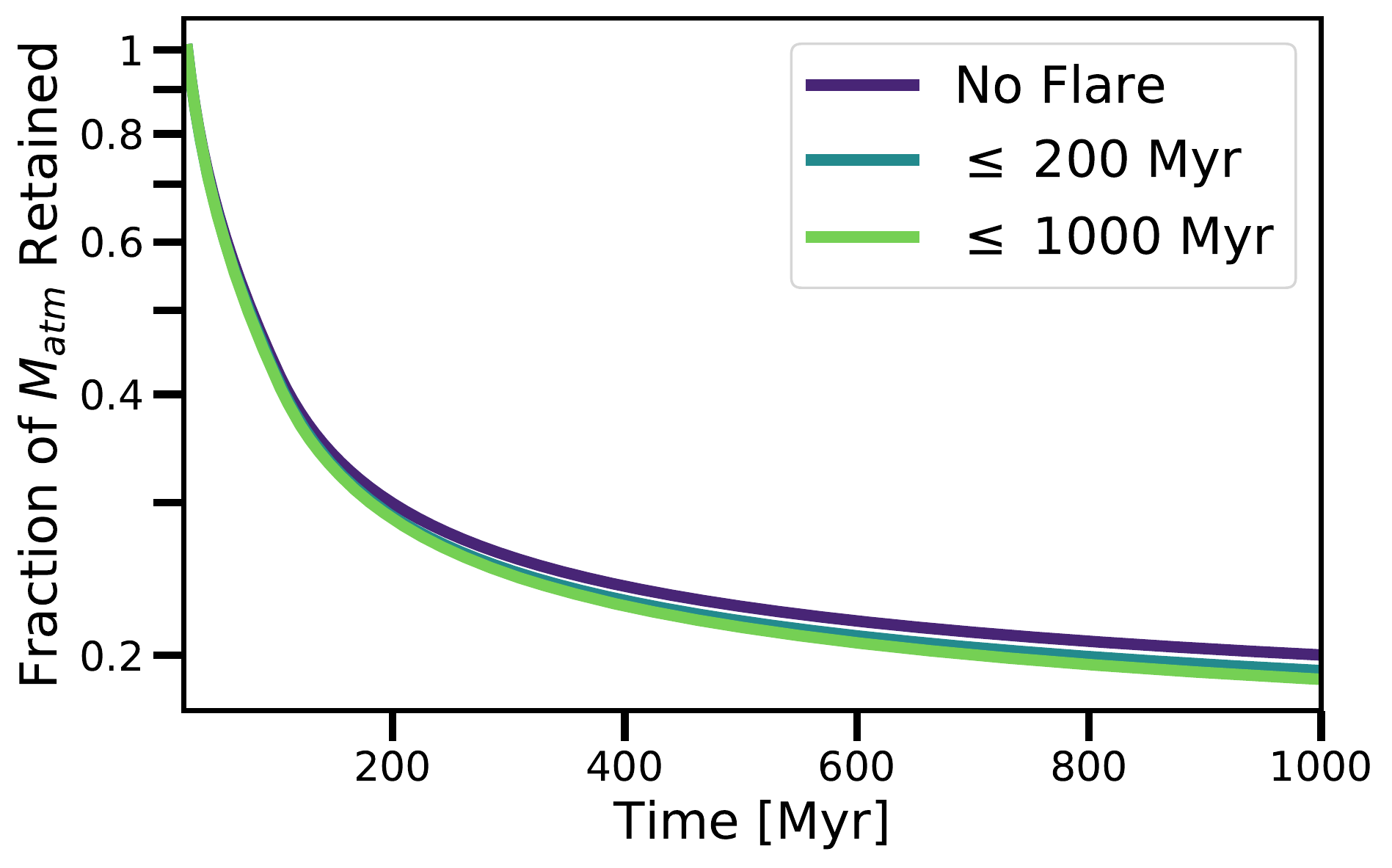}
\caption{Combining the methods of \cite{owen17} and \cite{owen20}, we calculated the atmospheric mass loss (bottom) for V1298\,Tau\,d. Purple lines correspond to a high-energy luminosity without flares; blue corresponds to flares being present in the first 200\,Myr; green corresponds to flares being present throughout all the 1000 Myr. The presence of flares affects the atmosphere mass, with flares present over a longer period of time removing more mass. \label{fig:mass_loss}}
\end{center}
\end{figure}

The evolution of the atmospheric mass as a function of time is shown in Figure~\ref{fig:mass_loss} for three different scenarios: one with no flares (purple, following the original methods of \cite{owen17} and \cite{owen20}), one with flares present for the first 200\,Myr (blue), and the third with flares present throughout 1000\,Myr (green). The injection of flares does show an effect on the final atmosphere mass. In the case of flares present for the first 200\,Myr, the planet loses 4\% more atmosphere than without flares. For the case of flares persistent across 1000\,Myr, the planet loses 7\% more atmosphere. Because the only parameter changing between models is $f(A)$ and we are evaluating ratios, the composition assumptions above listed do not have an effect on these values. 

As the mass-loss rate is linearly proportional the high-energy luminosity ($\dot{M} \propto f(A)$), we would expect the mass loss to increase with respect to flare amplitude. Thus stars that have flares that are 40-60$\times$ the base flux \citep[e.g.][]{paudel18, paudel19} will increase the mass loss proportionally, thus resulting in even smaller fractional retention of the atmospheric mass. Additionally, the radius of the planet changes due to atmospheric loss changes, originally starting at $R_p = 6.41R_\oplus$ and predicted to evolve to $R_{p, none} = 2.11 R_\oplus$. With the inclusion of flares, we find the final evolved radius to be $R_{p, 200} = 2.10 R_\oplus$ and $R_{p, 1000} = 2.09 R_\oplus$ for each model. Although these radius changes are smaller than the typical uncertainties in planet radii, these are physical changes brought about when including flares in atmospheric photoevaporation modeling.

The magnetic field of the planet should also be accounted for. \cite{kay16} explored the impacts of M dwarf coronal-mass ejections, sometimes associated with flares, on exoplanets and found that rocky exoplanets would need to generate magnetic field strengths of a tens to hundreds of Gauss, much stronger than fields in the Solar System, to protect its atmosphere against these events. The presence of a magnetosphere will directly affect the amount of energy from the flare that reaches the atmosphere, which then directly affects the mass loss efficiency, $\eta$. Future analyses should consider flares when thinking about mass loss efficiencies.

\subsection{Limitations \& Future Work}

Due to random alignments of stars with respect to our line of sight, one may a expect a handful of stars without measured $P_{\rm rot}$ which exhibit similar flare rates to those with measured $P_{\rm rot}$. These could be the result of either highly longitudinally symmetric spot distributions or a nearly pole-on stellar inclination, where polar starspots are consistently observed and thus show no obvious spot modulation. Additional follow-up of our sample with $v\sin{i}$ measurements could help to estimate the otherwise-difficult-to-measure latitudinal distribution of starspots. The presence of high-latitude spots could explain the lack of flare phase dependence (Figure~\ref{fig:flare_phase}) as polar spots are always visible. Sharp spectral lines in pole-on stars would make them amenable to direct measurement of starspot emission even for small coverage fractions of starspots since high spectral resolution can perceive individual weak photospheric lines arising from collective starspot emission.  Candidate pole-on stars in this sample would have to be reassessed to obtain $P_{\rm rot}$, since the measurement criteria discussed in Section~\ref{subsec:rots} may have missed potentially weak rotation signatures.

There are a few instances where \stella\ improperly characterizes a light curve feature as a flare. First, these improper characterizations are mainly constrained to stars with measurable rotation periods of $\leq 0.5$ days, contact binaries with sharp modulation features, and generally noisy light curves. This may be due to a limited sample of such stars in the original training set of the CNN. Second, the metric results of our CNN could be improved by re-sampling the training set to over-sample the lowest-amplitude flares in each batch set. As such, we hope to improve the overall recovery rate of flares with \stella\ by incorporating the new flare and non-flare examples identified in this work into the training, testing, and validation sets for future work. Third, users of \stella\ who want the most complete flare-only catalog with minimal contamination from noise should explore using a higher threshold probability metric. If users want a small sample of high fidelity flares, they should explore different threshold values to obtain the desired flare results. This metric can be set in the \stella\ code and will not affect the included pre-trained models.

The methods explored in this work provide hope for studying other events around young active stars, such as searching for exoplanet transits. Although this sample used real flares as the training set for the CNN, transits can be easily modeled and injected into young stellar light curves. CNNs have been trained to find transits in \ktwo\ data \citep[e.g.][]{pearson18}, however these searches have not been tailored towards young stars in \tess. We believe a similar CNN architecture to that created here would work well for transits and plan to explore this question in the near future.

The light curves processed in this work as well as the CNN models created will be hosted as a high-level science product (HLSP) at the Mikulski Archive for Space Telescopes (MAST). Additionally, \stella\ is an open-source package that can be downloaded through GitHub or via the Python Package Index. \stella\ includes the ability to create custom CNN model architectures, as well as all of the rotation measurements and flare fitting described in this work. As the models were trained on \tess\ two-minute data, users can predict flares in their own light curves of the same format with our models.\footnote{\url{http://adina.feinste.in/stella}} 

\section{Conclusions} \label{sec:conclusion}

We have presented the use of a convolutional neural network to identify flares and understand flare rates across a range of young ($t_{\rm age} <$ 800 Myr) ages in \tess\ two-minute light curves. Our key findings are the following:

\begin{enumerate}
    \item CNNs are a promising method for flare detection and are able to recover more flares at lower energies ($< 5\%$ flare amplitude) than previously used light curve detrending and sigma-outlier methods.
    
    \item There appears to be no preference for flare occurrence and spot phase across all flare energies. As flares and spots are both magnetically-driven events, this may be used to infer that neither hemisphere is spot-free, but rather the variability is driven by a different spot coverage fraction.
    
    \item Low-mass stars ($T_{\rm eff} < 4000$ K) show higher flare rates and larger flare energies across all ages of our sample. Hotter stars show higher flare rates and larger flare amplitudes at $t_{\rm age} \leq 50$ Myr and evolve to slow rates and weaker flares as they age. Stars with $T_{\rm eff} = 6200-15000$ show no flares at any amplitude for ages $t_{\rm age} > 50$ Myr.
    
    \item We find that across all ages, stars with $T_{\rm eff} \leq 4000$ K exhibit higher flare rates and flare energies than hotter stars. There is a slight increase in flare rate for cool stars in age bins $t_{\rm age} =$ 1-20\,Myr to 45-125\,Myr. Low mass stars are believed to reach the zero age main sequence at 50\,Myr \citep{Zuckerman04}. The peak of this flare rate in the 45-125\,Myr bin may indicate the turn on of these stars to the main sequence. However, a larger study of cool stars may be needed to further explore this relation.

    \item The largest flares were all found on late M dwarfs ($T_{\rm eff} \leq 3200$\,K). As the fully convective boundary is at $T_{\rm eff} \sim 3300$ K \citep{dorman89}, this suggests different internal processes driving the creation of such energetic flares.
    
    \item Modeled atmosphere mass loss due to photoevaporation suggests the inclusion of flares decreases the final atmospheric mass and planet radius when compared to models not accounting for flares. Thus, flares should be accounted for in such models moving forward.
    
    \item We hope to improve our CNN by incorporating our newly found flares and more examples of very fast ($< 0.5$ days) rotators in the training, validation, and test sets. We also plan on testing this CNN architecture and sliding box method to find new transiting exoplanets around young stars.

\end{enumerate}

\acknowledgements

We thank Dorian Abbot, Thaddeus Komacek, and James Owen for useful conversations, direction, and comments. We also thank Rodrigo Luger for his infinite Travis knowledge and skills. We thank our anonymous referee for valuable comments which improved this work.

This material is based upon work supported by the National Science Foundation Graduate Research Fellowship Program under Grant No. (DGE-1746045). Any opinions, findings, and conclusions or recommendations expressed in this material are those of the author(s) and do not necessarily reflect the views of the National Science Foundation. 
This work was funded in part through the NASA \tess\ Guest Investigator Program, as a part of Program G011237 (PI Montet).

This paper includes data collected by the \tess\ mission, which are publicly available from the Mikulski Archive for Space Telescopes (MAST). Funding for the TESS mission is provided by NASA’s Science Mission directorate. Many of the targets used in this study were part of the following \tess\ Guest Investigator programs (in numerical order): G011025, G011048, G011060, G011083, G011112, G011113, G011114, G011127, G011129, G011132, G011148, G011153, G011154, G011155, G011164, G011175, G011176, G011177, G011179, G011180, G011183, G011184, G011185, G011187, G011190, G011198, G011203, G011204, G011208, G011209, G011211, G011214, G011224, G011227, G011230, G011231, G011238, G011241, G011242, G011250, G011257, G011259, G011264, G011265, G011266, G011268, G011270, G011278, G011279, G011281, G011284, G011286, G011288, and G011294.

This project was developed in part at the Expanding the Science of \tess\ meeting, which took place in 2020 February at the University of Sydney.

This work is supported by the Deep Skies Community (\url{deepskieslab.com}), which helped to bring together the authors. 
Members of the Deep Skies Community have committed themselves to performing work in an equitable, inclusive, and just environment, and we hold ourselves accountable, believing that the best science is contingent on a good research environment. This manuscript has been authored by Fermi Research Alliance, LLC under Contract No. DE-AC02-07CH11359 with the U.S. Department of Energy, Office of Science, Office of High Energy Physics.

ADF would like to dedicate this paper to her grandmother, Diana Gross, for her endless support.

\software{%
    numpy \citep{numpy},
    matplotlib \citep{matplotlib},
    scipy \citep{jones01}
    lightkurve \footnote{\url{https://doi.org/10.5281/zenodo.2557026}},
    astropy \citep{astropy:2013, astropy18},
    \stella,
    tensorflow \citep{Abadi2016},
    \banyan \citep{gagne18},
    astroquery \citep{astroquery19}
    }

\facility{TESS}

\begin{deluxetable*}{l r r r r}[!ht]
\tabletypesize{\footnotesize}
\tablecaption{Number of Stars from Each Young Stellar Population \label{tab:grouppaarameters}}
\tablehead{
\colhead{Population} & \colhead{Age [Myr]} & \colhead{Count} & \colhead{Age Ref.}
}
\startdata
AB Doradus & 149$^{+51}_{-19}$ & 283 & \cite{Bell15}\\
Argus & 40-50 & 38 & \cite{zuckerman19}\\
$\beta$ Pictoris & 24 $\pm$ 3 & 144 & \cite{Bell15}\\
Carina & 45$^{+11}_{-7}$ & 78 & \cite{Bell15}\\
Carina-Near & $\sim$200 & 57 & \cite{Zuckerman06}\\
Columba & 42$^{+6}_{-4}$ & 129 & \cite{Bell15}\\
$\epsilon$ Chamaeleontis & 3.7 $^{+4.6}_{-1.4}$ & 14 & \cite{murphy13}\\
$\eta$ Chamaeleontis & 11 $\pm$ 3 & 3 & \cite{Bell15}\\
Hyades & 750 $\pm$ 100 & 61 & \cite{brandt15}\\
IC2391 & 50 $\pm$ 5 & 7 & \cite{barrado04}\\
IC2602 & 46 $^{+6}_{-5}$ & 22 & \cite{dobbie10}\\
Lower Centaurus Crux & 15 $\pm$ 3 & 338 & \cite{pecaut16}\\
Octans & 35 $\pm$ 5 & 80 & \cite{murphy15}\\
Platais 8 & $\sim$60 & 13 & \cite{platais98}\\
Pleiades & 112$\pm$5 & 11 & \cite{dahm15}\\
Taurus & 1-2 & 11 & \cite{kenyon95}\\
Tucana-Horologium & 45 $\pm$ 5 & 182 & \cite{Bell15}\\
TW Hydrae & 10 $\pm$ 3 & 33 & \cite{Bell15}\\
Upper Centaurus Lupus & 16 $\pm$ 2 & 436 & \cite{pecaut16}\\
Upper CrA & 10 & 6 & \cite{gagne18}\\
Upper Scorpius & 10 $\pm$ 3 & 13 & \cite{pecaut16}\\
Ursa Major cluster & 414 $\pm$ 23 & 3 & \cite{jones15}\\
$\chi$ For & $\sim$500 & 9 & \cite{pohnl10}
\enddata
\end{deluxetable*}

\end{document}